\documentclass[conference,compsoc]{IEEEtran}

\usepackage{tikz}
\usepackage{amsmath}

 \usepackage[normalem]{ulem}
 \usepackage{xspace}
 \usepackage{xcolor}
\newcommand{\ts}[3][]{#3}
\usepackage{comment}
\usepackage{graphicx}
\usepackage{adjustbox}
\usepackage{tikz}
\usepackage{pifont}
\usepackage{colortbl}
\usepackage{bm}
\usepackage{enumitem}
\usepackage{mdframed}
\usepackage{tcolorbox}
\usepackage{titlesec}
\usepackage{tabularx}
\usepackage{tabulary}
\usepackage{array}
\usepackage{setspace}
\allowdisplaybreaks

\newcommand{\sys}{PILOT\xspace}

\usepackage{listings}

\usepackage{titlesec}
\usepackage{hyperref}
\usepackage{amsmath}
\usepackage{algorithmic}

\usepackage{tcolorbox}
\usepackage{refcount}

\usepackage{amssymb}
\usepackage{pifont}
\usepackage{subcaption}

\usepackage[table]{xcolor} 
\usepackage{comment}
\usepackage{graphicx}
\usepackage{adjustbox}
\usepackage{tikz}
\usepackage{pifont}
\usepackage{colortbl}
\usepackage{bm}
\usepackage{enumitem}
\usepackage{mdframed}
\usepackage{tcolorbox}
\usepackage{tabularx}
\usepackage{tabulary}
\usepackage{array}
\usepackage{setspace}
\usepackage{tabularx}
\newcolumntype{R}{>{\raggedleft\arraybackslash}X}

\allowdisplaybreaks

\usepackage{hyperref}
\usepackage{amsmath}

\usepackage{tcolorbox}
\usepackage{refcount}

\newcounter{finding}
\newcommand{\findingbox}[1]{
  \stepcounter{finding}
  \begin{tcolorbox}[boxrule=0.2mm, colframe=black, top=2pt, bottom=2pt, left=3pt, right=3pt]
    \textbf{Result \thefinding.} #1
  \end{tcolorbox}
}

\usepackage{amssymb}
\usepackage{pifont}
\usepackage{subcaption}

\usepackage{xcolor}
\usepackage{multirow}
\usepackage{graphicx}
\usepackage[table]{xcolor} 

\usepackage{xspace}
\usepackage{pgfmath}
\usepackage{enumitem}

\usepackage{algorithm}
\usepackage{tikz}
\usepackage{pifont}

\usepackage{booktabs} 

\usepackage{eso-pic}
\AddToShipoutPictureBG*{%
  \AtPageLowerLeft{%
    \raisebox{1.2cm}{\hspace{0.75in}\parbox{0.85\paperwidth}{\footnotesize
    \copyright~2026, Momoko Shiraishi. Under license to IEEE.\\
    DOI 10.1109/SP63933.2026.00211}}}}

\begin{document}

\date{}


\title{PILOT: Command-line Interface Fuzzing via Path-Guided, \\Iterative Large Language Model Prompting} 


\author{
{\rm Momoko Shiraishi}\\
The University of Tokyo\\
shiraishi@os.is.s.u-tokyo.ac.jp
\and
{\rm Yinzhi Cao}\\
Johns Hopkins University\\
yinzhi.cao@jhu.edu
\and
{\rm Takahiro Shinagawa}\\
The University of Tokyo\\
shina@is.s.u-tokyo.ac.jp
} 

\maketitle

\begin{abstract}
Command-line interface (CLI) fuzzing tests programs by mutating both command-line options and input file contents, thus enabling discovery of vulnerabilities that only manifest under specific option-input combinations. Prior works of CLI fuzzing face the challenges of generating semantics-rich option strings and input files, which cannot reach deeply embedded target functions. 
This often leads to a \textcolor{black}{missed detection} of such a deep vulnerability using existing CLI fuzzing techniques. 

In this paper, we design a novel Path-guided, Iterative LLM-Orchestrated Testing framework, called \sys, \textcolor{black}{to generate effective initial seeds for fuzzing CLI applications}. 
The key insight is to provide potential call paths to target functions as context to LLM so that it can better generate CLI option strings and input files. Then, \sys iteratively repeats the process, and provides reached functions as additional context so that target functions are reached. 

Our evaluation on real-world CLI applications demonstrates that \sys, \textcolor{black}{combined with existing fuzzers}, achieves higher coverage than state-of-the-art fuzzing approaches and discovers 51 zero-day vulnerabilities.
We responsibly disclosed all the vulnerabilities to their developers and so far 41 have been confirmed by their developers with 33 being fixed and \textcolor{black}{five} assigned CVE identifiers.


\end{abstract}

\section{Introduction}

Fuzzing is a well-known and widely adopted approach to systematically discover software bugs and vulnerabilities, e.g., in Command-Line Interface (CLI) programs, with unexpected, malformed, or random inputs.  Traditional fuzzing techniques, according to a survey of 102 recent fuzzing papers~\cite{lee2025zigzagfuzz}, often operate with fixed option configurations, making them challenging to reach deeply-embedded vulnerable functions that are only triggerable with certain option strings. 
 Even recent directed fuzzing~\cite{huang2022beacon, luo2023selectfuzz}, which use distance metrics to guide mutations toward target functions, may face similar challenges because without a correct option combination, distance metric may stay at infinity regardless of other mutations.

Therefore, an important problem that CLI fuzzing approaches need to address is the generation of different combinations of option strings.  Traditional CLI fuzzing approaches~\cite{wang2023carpetfuzz,song2020crfuzz,zhang2023fuzzing,lee2025zigzagfuzz} rely on CLI option parsers, which only capture syntactic knowledge but not semantic relationships between different options.  ProphetFuzz~\cite{wang2024prophetfuzz}, the only, state-of-the-art LLM-based CLI fuzzing approach, parses the man pages of CLI options using LLM to generate CLI option combinations that are likely to trigger vulnerabilities. 
%
%
%
 However, sometimes, man pages do not contain the deep knowledge of different option combination either.

 More importantly, one unaddressed research challenge---beyond the aforementioned option string combination---is the generation of input files.  Specifically, input files need to satisfy certain structural properties, which include but are not limited to proper file headers, valid codec parameters, correct container formats, and well-formed data structures.   However, to the best of our knowledge, such input files are mostly provided beforehand as prior knowledge with manual efforts in state-of-the-art approaches. This not only brings additional human efforts, but also leads to challenges in triggering certain parts of CLI tools, which require a specific format.

In this paper, we design a novel Path-guided, Iterative LLM-Orchestrated Testing (\sys) framework \textcolor{black}{to generate effective initial seeds for fuzzing CLI applications}. 
The key idea is that \sys identifies potential call paths to target functions and incorporates these paths into prompts as context so that \sys can have a better understanding of the command line specification, thus generating different combinations of CLI option strings. Meanwhile, \sys automatically discovers and orchestrates native command-line tools to generate semantically valid input files that enable reaching the target functions.
 The entire process is iterative. If the inputs from LLMs cannot reach the target function, \sys provides feedback indicating which functions along the candidate paths are covered and then re-queries LLM for another set of potential option combinations and input files.

We implemented a prototype of \sys and plan to release it at 
\url{https://github.com/momo-trip/PILOT}.
 Our evaluation across 43 widely used programs shows that \sys discovered 51 zero-day vulnerabilities.  We responsibly disclosed all of them to their developers and so far, 41 have been confirmed by developers with 33 being fixed and \textcolor{black}{five} assigned CVE identifiers.  We also compared \sys with state-of-the-art CLI fuzzing approaches, and the results showed 
 \textcolor{black}{that combining PILOT's seeds with existing fuzzers yields substantial improvements in code coverage,
 i.e. 20\% over CarpetFuzz~\cite{wang2023carpetfuzz}, 28\% over ProphetFuzz~\cite{wang2024prophetfuzz}, 15\% over ZigZagFuzz~\cite{lee2025zigzagfuzz}, and 66\% over SelectFuzz~\cite{luo2023selectfuzz}. 
 }

\section{Overview}\label{sec:back}
In this section, we first describe a motivating example of seed generation for CLI fuzzing, and then give an overview of our solution.

\subsection{A Motivating Example}
We describe a motivating example with a zero-day vulnerability found by \sys in FFmpeg~\cite{ffmpeg}, a multimedia framework for video and audio processing, which is widely used in popular applications
 like web browsers, media servers, and video conferencing systems.  The vulnerability is 
 a zero-day buffer overflow due to a negative size parameter, located in FFmpeg's Real-time Transport Protocol (RTP) streaming functionality of mp4 files~\cite{ffmpeg_protocols}. 
 It can be triggered either locally via command line or remotely through network streaming protocols, leading to arbitrary code execution through controlled heap/stack corruption.  We have responsibly disclosed this vulnerability to the FFmpeg developers, who have acknowledged and fixed the vulnerability. 
\textcolor{black}{This vulnerability has been assigned CVE-2026-30754.}



Figure~\ref{fig:mot_example} shows a simplified version of the vulnerable code.  Line 1 shows the command line that triggers the vulnerability.  Specifically, the sample size table (stsz) atom in the crafted MP4 file (``malicious.mp4'') in a valid H.264 format specifies a sample size with trailing byte for exploitation. 
 The root cause of the vulnerability is that FFmpeg does not properly validate Network Abstraction Layer (NAL) unit sizes when processing H.264/HEVC streams for RTP transmission, leading to a negative size parameter passed to \texttt{memcpy} at Line 44. 
Specifically, the vulnerability occurs when the stsz, i.e., each frame's byte size, declares a frame size with 1-3 trailing bytes beyond the complete NAL unit.
 After a complete NAL unit is processed, 1 trailing byte, for example, remains and the loop at Line 22 continues. At Line 25, \texttt{ff\_nal\_mp4\_find\_startcode()} returns NULL since \texttt{end - r < 4}. The code sets \texttt{r1 = end} at Line 26, then executes \texttt{r += 4} at Line 27, causing \texttt{r = end + 3}. The calculation \texttt{len = r1 - r = -3} at Line 32 produces a negative value passed to \texttt{nal\_send()} without validation. 
At Line 44, \texttt{memcpy(s->buf\_ptr, buf, -3)} casts -3 to \texttt{size\_t}, becoming \texttt{0xFFFFFFFFFFFFFFFD} (18 exabytes). Given a typical 1500-byte RTP buffer, this attempts to copy 12 trillion times more data than the buffer can hold, resulting in \textcolor{black}{a buffer overflow}. 

\begin{figure}[!htbp]
\begin{lstlisting}[language=C, basicstyle=\fontsize{6.5pt}{7.5pt}\ttfamily]
    // exploit code: ./ffmpeg -i malicious.mp4 -c copy -f rtp rtp://127.0.0.1:1234
    // ffmpeg.c - Main entry point
    int main(int argc, char **argv) {
       // ... initialization ...
       ret = transcode();  // Start transcoding
       // the call chain is transcode -> muxer_thread -> av_interleaved_write_frame -> rtp_write_packet -> ff_rtp_send_h264_hevc -> nal_send
       return ret;
    }
    // rtpenc.c - RTP encoder
    static int rtp_write_packet(AVFormatContext *s1, AVPacket *pkt) {
       // ... RTP packet setup ...
       if (st->codecpar->codec_id == AV_CODEC_ID_H264 ||
           st->codecpar->codec_id == AV_CODEC_ID_HEVC) {
          ff_rtp_send_h264_hevc(s1, pkt->data, pkt->size);  // Process H.264/HEVC
       }
       return 0;
    }    
    // rtpenc_h264_hevc.c - H.264/HEVC RTP packetizer
    void ff_rtp_send_h264_hevc(AVFormatContext *s1, const uint8_t *buf1, int size) {
       const uint8_t *r = buf1, *end = buf1 + size;
       RTPMuxContext *s = s1->priv_data; 
       while (r < end) {
          const uint8_t *r1;        
          if (s->nal_length_size) {  // MP4 format
             r1 = ff_nal_mp4_find_startcode(r, end, s->nal_length_size);
             if (!r1) r1 = end;
             r += s->nal_length_size;  // VULNERABLE: No boundary check
          } else {
             while (r < end && *r == 0) r++;
             r1 = ff_nal_find_startcode(r, end);
          }     
          int len = r1 - r;  // VULNERABLE: Can be negative
          nal_send(s1, r, len, r1 == end);
          r = r1;
       }
    }
    static void nal_send(AVFormatContext *s1, const uint8_t *buf, int size, int last) {
       RTPMuxContext *s = s1->priv_data;
       // No validation of size parameter!
       av_log(s1, AV_LOG_DEBUG, "Sending NAL %x of len %d\n", 
              buf[0] & 0x1F, size);
       if (size <= s->max_payload_size) {
          // ...
          memcpy(s->buf_ptr, buf, size);  // CRASH: size = -3
          // Cast to size_t = 0xFFFFFFFFFFFFFFFD
          s->buf_ptr += size;
       }
    }
\end{lstlisting}
\vspace{-0.15in}
\caption{A motivating example illustrating a zero-day buffer overflow vulnerability found by \sys. The code is slightly changed and simplified for better presentation.}
\label{fig:mot_example}
\end{figure}


\subsection{Research Challenge and Overall Solution}

The main research challenge is how to reach the vulnerable location with the correct inputs that can trigger the vulnerability.  There are two conditions for this specific motivating example in Fig.~\ref{fig:mot_example}: (i) a valid input file, i.e., a H.264/HEVC input file with proper NAL unit structures, and (ii) a correct combination of option strings, which are ``-i file -c copy -f rtp address'' for this vulnerability.
 More specifically, the input file must be valid in terms of containing valid H.264/HEVC codec identification to pass the condition \texttt{st->codecpar->codec\_id == AV\_CODEC\_ID\_H264} (at Line 12) and having properly initialized codec parameters to survive \texttt{av\_interleaved\_write\_frame} processing.

To the best of our knowledge, state-of-the-art CLI fuzzing approaches~\cite{wang2023carpetfuzz, wang2024prophetfuzz, lee2025zigzagfuzz} have difficulties in triggering this vulnerability, because they cannot generate the seeds that satisfy or are close to the aforementioned two conditions.

\noindent{\textbf{Valid Input File Generation.}}
 First, prior works mainly adopt a manually-crafted input file or one from the test cases, which is labor-intensive and may not capture the complex condition, e.g., satisfying the  H.264/HEVC codec identification.

\noindent{\textbf{Semantic Option Combination.}}
Second, they mainly adopt some semantic understanding methods, e.g, analyzing the source code's option parser (such as \texttt{getopt}), man pages (like \texttt{man ffmpeg}), or options from help output (such as \texttt{./ffmpeg --help}).  None of them defines the behaviors of different option combinations like ``-i file -c copy -f rtp address'', thus failing to generate such option strings.


\sys solves the problem to satisfy the two conditions by directed fuzzing using path-guided Large Language Model (LLM) prompting. Let us use \texttt{ffmpeg} as an example for illustration.  First, 
 \sys queries an LLM to identify and install \texttt{ffmpeg} and generate base input files with proper codec parameters and container structures (e.g., via \texttt{ffmpeg -f lavfi -i testsrc -c:v libx264 output.mp4}). 
 Note that the tool used for file generation may coincide with the fuzzing target, as in this example, but is generally a separate utility.
 This approach ensures deep semantic validity—files generated by actual encoders contain valid codec parameters and NAL unit structures, where subsequent mutations can introduce targeted malformations to trigger vulnerabilities. Second, \sys leverages path-guided exploration and provides the LLM with the call chain from main to potential targets. 
 This path-guided prompting enables the LLM to systematically examine code at each step, discovering that: (1) \texttt{print\_sdp} is triggered by the \texttt{-sdp\_file} option, (2) SDP generation works in conjunction with RTP output requiring \texttt{-f rtp}, and (3) HEVC codec input is needed to reach the HEVC-specific path through \texttt{extradata2psets\_hevc}.

Note that the process is an iterative approach.  \sys executes each generated command, measures coverage to verify target function reachability, and provides feedback for iterative refinement when commands fail or the target function is not reached. 
 By doing so, \sys successfully generates validated initial seeds that reach deep program functionality.
Again, we use the FFmpeg case as an example. 
 After \sys targets the high-centrality function \texttt{ff\_isom\_write\_hvcc()} to explore RTP-related subsystems, \sys further discovers the \texttt{nal\_send()} function, which contains the vulnerability in \textcolor{black}{our} motivating example.

\section{Design} 
In this section, we describe the design of \sys from its system architecture and then present the detailed core techniques of \sys.

%

\begin{figure}[!t]
    \centering
    \includegraphics[width=\linewidth]{pilot_overview.pdf}
    \caption{\textcolor{black}{The design overview of \sys.}}
    \label{fig:design}
\end{figure}


\newcommand{\circledzero}{%
  \tikz[baseline=(char.base)]{
    \node[shape=circle,draw,inner sep=0.6pt,minimum size=7pt,line width=0.2pt] (char) {\scriptsize 0};
  }%
}


\subsection{Design Overview}

\newcounter{stepscounter}
\newlist{steps}{enumerate}{1}
\setlist[steps]{label=\textbf{Step \arabic*}.}
\sys leverages LLMs to generate effective initial seeds for CLI fuzzing, consisting of command-line strings and input files, which are later used by state-of-the-art CLI fuzzers.
To achieve this, \sys performs two key functions.
First, it coordinates seed generation by iteratively communicating with the LLM, sending tailored prompts together with supplemental information such as the target function, potential call path to the function, and coverage feedback from the previous seed.
Second, it provides an API that allows the LLM to interact with a sandboxed shell environment, where it can read target source-code fragments, write and execute a shell script, and install necessary command packages to generate input files.
These capabilities enable the LLM to flexibly generate files and obtain contextual information as needed, facilitating iterative refinement of command-line seeds.

\autoref{fig:design} illustrates the overall design of \sys.
First, \sys instructs the LLM to initialize the sandboxed shell environment for input file generation (\autoref{fig:design} (a), \autoref{subsec:command_install}).
The LLM generates a script that installs the necessary command packages for input file generation, which is essential for producing valid and diverse input files.
Next, \sys selects a set of target functions (\autoref{fig:design} (b), \autoref{subsec:function_selection}).
Because LLM inference is costly, \sys prioritizes target functions based on centrality metrics to achieve high coverage efficiently.

\sys then prompts the LLM to generate command-line options intended to reach the selected target function (\autoref{fig:design} (c), \autoref{subsec:seed_generation}).
To facilitate this, potential call paths are precomputed through static source-code analysis and provided to the LLM as contextual information to help it understand how the target function can be reached within the program.
The LLM reads relevant source-code fragments as needed (\autoref{fig:design} (d)) and returns a shell script to \sys that includes candidate command-line strings and commands to generate input files designed to reach the target function.

\sys executes the generated commands, collects coverage data, and, if the target function is not reached, prompts the LLM to refine the command line using the obtained coverage feedback.
This iterative seed generation process is repeated up to a fixed number of attempts, and if the LLM still fails to produce a command line that reaches the target function, \sys moves on to the next target function to improve overall coverage efficiently.
The following subsections describe the details of the core techniques of \sys in accordance with the procedure outlined in \autoref{alg:testgen}.

\subsection{Initialization for Input File Generation}
\label{subsec:command_install}

Generating semantically valid input files is essential for effective CLI fuzzing. Rather than relying on random byte mutations that often fail parsing checks, \sys automatically identifies and configures command-line tools that produce well-formed input files ($\circledzero$ of \autoref{alg:testgen}).

\sys queries the LLM to discover standard command-line tools for creating valid inputs for a target command. 
To prevent trivial or ad-hoc file generation (e.g., \texttt{printf "\textbackslash x89PNG..."\ >\ image.png} or \texttt{echo "data"\ >\ video.mp4}),
\sys employs carefully designed prompts that explicitly require well-established, standard command-line tools.
For example, when fuzzing \texttt{ffmpeg}, \sys asks the LLM to identify standard command-line tools that can produce valid file formats, such as \texttt{ffmpeg} itself, \texttt{convert} (ImageMagick) or \texttt{gstreamer}.
The LLM then provides installation commands for the target OS.
After executing the installation commands, \sys verifies that each tool is correctly installed and accessible by checking command availability (using \texttt{command -v}). If a tool fails verification, \sys repeats the installation and verification process. This iterative approach continues until the tool is successfully installed and verified.


\subsection{Target Function Selection}
\label{subsec:function_selection}

While \sys can potentially target any function in a program given appropriate prompts and context, practical constraints necessitate strategic selection of which functions to prioritize. Each seed generation attempt consumes LLM tokens, and the token budget is finite. Therefore, \sys's primary objective is to maximize code coverage within this given token budget;
\sys must carefully select which functions to target to achieve the broadest coverage most efficiently.

Different programs exhibit distinct structural characteristics in their function call graphs. 
These structural properties can predict which functions will yield better coverage when targeted.
Therefore, \sys leverages graph centrality metrics, which identify structurally important functions in the call graph, to guide target selection. 
Specifically, \sys employs an {\it adaptive target selection mechanism}.
\sys analyzes each program's call graph structure (\autoref{alg:testgen} \ding{172}) and applies learned decision rules (detailed in Section~\ref{ablation}) to automatically 
select the \textcolor{black}{most effective} centrality-based strategy for that program (\autoref{alg:testgen} \ding{173}) . 
This adaptive approach maximizes coverage within the available token budget by selecting the strategy most suited to each program's unique structural characteristics.

\begin{algorithm}[!t]
\caption{Procedure of \sys} 
\footnotesize
\label{alg:testgen} 
\begin{algorithmic}[1]
\REQUIRE Program source code, set of functions $F = \{f_1, f_2, ..., f_n\}$
\ENSURE Set of commands

\STATE $\text{NativeToolConfiguration}(commands)$ \COMMENT{\circledzero\ Environment setup} 
\STATE Construct call graph $G = (V, E)$ where $V$ is the set of functions and $E$ represents function calls
\STATE $U \leftarrow F$ \COMMENT{$U$: Set of unvisited functions, initially all functions}
\STATE \textcolor{black}{$T \leftarrow \emptyset$ \COMMENT{Set of already-targeted functions}}
\STATE $N_{trial} \leftarrow$ Maximum number of attempts per function
\STATE $N_{target} \leftarrow$ Maximum number of functions to target
\STATE \textcolor{black}{$N_{branch} \leftarrow$ Maximum number of branch refinement attempts}
\STATE \textcolor{black}{$n_{processed} \leftarrow 0$ \COMMENT{Initialize processed counter}}
\STATE $\mathbf{c} \leftarrow$ Centrality score vector \COMMENT{\ding{172} Initial function scoring} 

\WHILE{ \textcolor{black}{$U \setminus T \neq \emptyset$ \AND $n_{processed} \leq N_{target}$}}
    \STATE \textcolor{black}{$f_{target} \leftarrow \arg\max_{f_i \in U \setminus T} \mathbf{c}[i]$ \COMMENT{\ding{173} Select function with highest centrality score}}
    \STATE \textcolor{black}{$T \leftarrow T \cup \{f_{target}\}$ \COMMENT{Mark as targeted}}
    \STATE $trial \leftarrow 0$ \COMMENT{Initialize attempts counter}
    \STATE $covered$ $\leftarrow$ false
    \STATE \textcolor{black}{$commands \leftarrow \text{GenerateCommand}(f_{target}, G)$ \COMMENT{\ding{174} LLM commands generation}}
    \WHILE{$trial \leq N_{trial}$ \AND not $covered$}
        \STATE $program\_state \leftarrow \text{ExecuteCommand}(commands)$ \COMMENT{\ding{175} Command execution}
        \STATE $cov \leftarrow \text{AnalyzeCoverage}(program\_state)$ \COMMENT{\ding{176} Coverage analysis}
        
        \IF{$f_{target} \in cov.covered\_functions$}
            \STATE $covered$ $\leftarrow$ true
        \ELSE
            \STATE $trial \leftarrow trial + 1$ \COMMENT{Increment attempts counter}
            
            \IF{$trial > N_{trial}$}
                \STATE \textbf{break}
            \ELSE
                \STATE $commands \leftarrow \text{Refine}(commands, cov, f_{target})$ \COMMENT{\ding{177} Refine the commands based on the feedback}
            \ENDIF
        \ENDIF
    \ENDWHILE
    \IF{\textcolor{black}{$covered$}}
        \STATE \textcolor{black}{$branch\_trial \leftarrow 0$ \COMMENT{Initialize branch attempts counter}}
        \WHILE{\textcolor{black}{$branch\_trial < N_{branch}$}}
            \STATE \textcolor{black}{$commands \leftarrow \text{RefineBranches}(f_{target}, cov)$ \COMMENT{\ding{178} Branch-level refinement}}
            \STATE \textcolor{black}{$program\_state \leftarrow \text{ExecuteCommand}(commands)$}
            \STATE \textcolor{black}{$cov' \leftarrow \text{AnalyzeCoverage}(program\_state)$}
            \IF{\textcolor{black}{$cov'.branches \setminus cov.branches \neq \emptyset$}}
                \STATE \textcolor{black}{$cov \leftarrow cov'$}
                \STATE \textcolor{black}{\textbf{break}}
            \ENDIF
            \STATE \textcolor{black}{$branch\_trial \leftarrow branch\_trial + 1$}
        \ENDWHILE
    \ENDIF
    \STATE $n_{processed} \leftarrow n_{processed} + 1$
    \STATE \textcolor{black}{$U \leftarrow U \setminus cov.covered\_functions$ \COMMENT{\ding{179} Remove newly covered functions}}
    \STATE $\mathbf{c} \leftarrow$ Recalculate updated centrality scores over $U$ in $G$ \COMMENT{\ding{172} Function scoring}
\ENDWHILE

\STATE $commands \leftarrow$ All generated commands
\STATE $seeds \leftarrow \text{ConvertToSeeds}(commands)$ \COMMENT{\ding{180} Transform commands into fuzzing initial seeds}
\STATE $fuzzing\_results \leftarrow \text{ExecuteFuzzing}(seeds)$ \COMMENT{\ding{181} Run fuzzing with generated initial seeds}
\RETURN $fuzzing\_results$
\end{algorithmic}
\end{algorithm}

\subsection{Path-guided Context Prompting}
\label{subsec:seed_generation}

Path-guided context prompting enhances LLM-based seed generation by providing explicit execution path information and enabling LLMs to understand the relationship between command-line arguments and target functions efficiently. 
This technique is complemented by an LLM-user interface that works together to overcome the inherent limitations of LLM context windows~\cite{window}.
The following subsection describes the path-guided context prompting and the interface that enables autonomous code exploration.

\subsubsection{Initial Prompt}
The initial phase leverages static analysis to identify potential execution paths from the main function to target functions. 
\textcolor{black}{Specifically, \sys constructs a whole-program call graph by parsing the AST of all source files using the Clang AST parser~\cite{libtooling}, extracting function references including both direct calls and function pointer assignments to build caller-callee relationships. 
\sys selects target functions only from those reachable from \texttt{main()} in the constructed graph; functions with no identified path are not selected as targets.}

~\autoref{fig:path-guided-prompt} shows an example prompt for exploring a target function (\texttt{ff\_isom\_write\_hvcc()}) in the \texttt{ffmpeg} command.
In the ``Function Call Relationship'' section (\ding{182} of ~\autoref{fig:path-guided-prompt}), multiple candidate execution paths from the main function to the target function are provided, each with their corresponding file locations and line numbers.
The file name and line number for each function represent where the function is defined, not where it is called from, in order to provide stable reference points and reduce complexity, as multiple invocation paths may exist for the same execution sequence.
For example, \texttt{main@ffmpeg.c:2932} indicates that the \texttt{main} function is located in the file \texttt{ffmpeg.c} at line \texttt{2932}.
The LLM is then instructed to generate a shell script (\texttt{run\_test.sh}) containing commands designed to trigger these execution sequences and reach the target function (\autoref{alg:testgen} \ding{174}).
Due to the large number of potential execution paths—sometimes approaching 100—LLM input token limitations occasionally prevent including all possible paths at once. To overcome this constraint, \sys exports the complete set of execution paths to an external file. This allows the LLM to access the full range of candidates through another read request-response chat (detailed in the subsequent \autoref{auto}), ensuring comprehensive analysis despite token restrictions.


\begin{figure}[!t]
\begin{tcolorbox}
[colback=blue!2,colframe=blue!60!black,title=Example Prompt for \{target command\}, fonttitle=\sffamily\bfseries]
\sffamily\footnotesize
\begin{flushleft}
\textcolor{blue!70!black}{\textbf{Task:}} Write ffmpeg test commands that reach the target function [ff\_isom\_write\_hvcc@ffmpeg.c:1084].\\
Please include commands to create valid input files using appropriate standard tools: \textbf{ffmpeg, convert, sox, mencoder}. Do not manually construct file headers or use placeholder data. Invalid files cause early rejection and low coverage.\\
$ $  \\
\textcolor{blue!70!black}{\textbf{Target Function Definition:}}\\
int ff\_isom\_write\_hvcc(AVIOContext *pb, ...) {...}\\
$ $  \\
\ding{182} \textcolor{red!65!black}{\textbf{Function Call Relationship from main() to the target function:}}\\
- Path candidate 1\\
main@ffmpeg.c:2932\\
$\rightarrow$ transcode/ffmpeg.c:2880\\
$\rightarrow$ transcode\_init@ffmpeg.c:2770\\
$\rightarrow$ init\_output\_stream@ffmpeg.c:2650\\
... \\
$\rightarrow$ extradata2psets\_hevc@sdp.c:226\\
$\rightarrow$ ff\_isom\_write\_hvcc@hevc.c:1084\\
$ $  \\
- Path candidate 2 ...\\
...\\
$ $  \\
\ding{183} \textcolor{teal!65!black}{\textbf{Response Format:}} Please respond using one of the following three modes:
\begin{enumerate}
\item read\_data: Request to read file contents
\item modify\_data: Update file contents
\item execute\_command: Run any command for testing
\end{enumerate}
\ding{184} \textcolor{orange!75!black}{\textbf{Directory Structure:}}\\
workspace/ \\
\hspace*{0.5cm}\{target\_program\}-xxxx/ \\
\hspace*{1.0cm}src/ \ding{185} \# Source files (file1.c, file2.c, ...) \\
\hspace*{1.0cm}run\_test.sh \ding{186} \# LLM response script \\
\hspace*{1.0cm}... \\
\hspace*{0.5cm}execute.sh \ding{187} \# Script for execute\_command mode
\end{flushleft}
\end{tcolorbox}
\caption{Example prompt for exploring a target function. The prompt directs the generation of a shell script file (\ding{186} \texttt{run\_test.sh}) that will trigger execution to reach the target function.}
\label{fig:path-guided-prompt}
\end{figure}

\subsubsection{Autonomous Code Exploration}~\label{auto}
Due to the limitations of LLMs' context windows, providing an entire codebase at once is often impractical. Instead, \sys implements an autonomous code exploration method that enables LLMs to independently navigate and select relevant files for coding tasks. 
To enable this autonomous exploration, \sys defines an API with the following three modes, requiring the LLM to select one in each response (described in \ding{183} of ~\autoref{fig:path-guided-prompt}).

\noindent{\bf{Read files.}} \hspace{0.05in}
The \texttt{read\_data} mode serves as the LLM's primary information-gathering mechanism. It allows the model to request specific file contents by providing the hierarchical target program directory structure (\ding{184} of ~\autoref{fig:path-guided-prompt}). 
This serves as a \textit{map} of the project and provides the interface to know the details of each source file (\ding{185}).
For larger files that exceed context window limitations, \sys supports targeted examination of specific line ranges. 
The LLM can proactively specify particular sections of interest, or \sys automatically provides guidance when token limits are reached. 
When the requested range exceeds the token limit, \sys displays guidance in a prompt, such as ``Exceeding token limit, content truncated. To view the complete content of [file\_path], please use read\_data mode and set file\_slice (specified range) to read each section separately.'' 
This interactive capability enables the LLM to progressively build understanding of the project structure and functionality through strategic file selection and guided exploration.

\noindent{\bf{Write files.}} \hspace{0.05in}
The \texttt{modify\_data} mode enables precise file modifications. The LLM can specify exact line ranges within files and provide replacement content, implementing targeted changes without needing to process entire files. This mode supports various modification patterns including complete replacements, deletions, and full file rewrites, giving the LLM fine-grained control over code changes. \textcolor{black}{In practice, \sys primarily uses this mode to write and update the test script (\texttt{run\_test.sh}) that contains the generated commands, as well as the auxiliary execution script (\texttt{execute.sh}) used by the \texttt{execute\_command} mode.}

\noindent{\bf{Execute commands.}} \hspace{0.05in}
The \texttt{execute\_command} mode completes the interaction loop by allowing the LLM to run shell commands through a shell script in the workspace (\ding{187} execute.sh) to test modifications or gather additional runtime information. By executing the script, the LLM can install necessary additional packages or extract information that might be challenging to parse from static source files. 

\subsection{Efficient Iterative Refinement}
\label{subsec:iterative_refinement}

\sys employs efficient iterative refinement techniques to maximize code coverage when LLMs struggle to reach target functions. 
\textcolor{black}{This process consists of four steps: covered-path feedback, target switching, branch-level refinement, and candidate function set update.}

\subsubsection{Covered-path Feedback}
If the target function is not reached (\autoref{alg:testgen} \ding{175} and \ding{176}), the feedback phase utilizes execution results to refine and guide subsequent seed generation attempts (\autoref{alg:testgen} \ding{177}). After executing generated commands, \sys analyzes which functions were actually covered during execution. This feedback of executed function list enables the LLM to understand what went wrong and adjust its seed generation strategy accordingly to reach the target function.

\subsubsection{Target Switching}
While targeting high-centrality functions is an efficient method to increase coverage, in practice, LLMs cannot always cover target functions successfully. To efficiently allocate tokens when dealing with difficult-to-cover functions, \sys implements a pragmatic approach: the LLM generates commands for each target function up to a configurable parameter $N_{trial}$.
If coverage remains unachieved after these attempts, \sys redirects resources by switching to the next highest-priority function in the queue (\autoref{alg:testgen} line 25).

Note that \sys implements switching based on the number of test attempts rather than time limits. This design choice stems from \sys's three distinct modes for request and response (described in~\autoref{auto}). The inclusion of a read file mode prevents premature transitions to subsequent target generation before completing the seed generation attempt. Therefore, the limit is set at the point where the writing test operation consistently succeeds for $N_{trial}$ consecutive runs.

\subsubsection{Covering Branches}
After successfully reaching a target function, \sys refines test generation to achieve finer-grained coverage at the branch level \textcolor{black}{(\autoref{alg:testgen} \ding{178})}. While function-level coverage provides a high-level view of program exploration, branch coverage reveals which execution paths within functions have been exercised. Once a function is covered, \sys systematically guides the LLM by providing detailed context about uncovered branches. Specifically, \sys includes in the prompt: (1) precise information about which branches remain uncovered, extracted through coverage analysis, (2) the target function definition, and (3) the directory structure to enable the LLM to comprehensively understand the codebase context. 
By explicitly supplying this structured information to the LLM, \sys enables the model to reason about how to trigger uncovered branches through adjustments to command-line arguments, input file contents, or environmental conditions. This refinement process is iterative: \sys repeats the generation-execution-analysis cycle until at least one new branch is discovered, or until \textcolor{black}{a maximum $N_{branch}$ refinement attempts have been made}. 

\textcolor{black}{
\subsubsection{Candidate Function Set Update}
After processing each target function, PILOT removes newly covered functions from the candidate set $U$ and recalculates centrality scores over the remaining candidates in $G$ (\autoref{alg:testgen} \ding{179}). 
This dynamic update ensures that subsequent target selection reflects the current coverage state, avoiding redundant exploration of already-covered regions.
}

\begin{figure}[t]
\begin{lstlisting}[language=bash,basicstyle=\footnotesize\ttfamily, numbers=none]
#!/bin/bash
# Commands for preparing input files
dd if=/dev/zero of=raw.yuv bs=1 count=12288
ffmpeg -f rawvideo -pix_fmt yuv420p -s 64x64 -r 1\
  -i raw.yuv -c:v libx265 -preset ultrafast \
  -x265-params keyint=1:aud=1 -frames:v 1 input.mp4

# The main test command
timeout 3s ./ffmpeg -re -i input.mp4 -c:v copy \
  -f rtp rtp://127.0.0.1:5030
\end{lstlisting}
\caption{\textcolor{black}{Example shell script generated by \sys for the FFmpeg RTP streaming target.}}
\label{fig:generated-script}
\end{figure}

\subsection{Initial Seed Extraction from LLM-generated Test Suites}
\label{subsec:seed-extraction}

The iterative exploration cycle described above produces multiple shell scripts (\texttt{run\_test1.sh}, \texttt{run\_test2.sh}, etc.) that execute the target program with different configurations. To use these tests as fuzzer seeds, \sys transforms them into the format required by the target fuzzer by extracting command-line option strings and collecting input files (\autoref{alg:testgen} \ding{180}).
\sys is designed to generate initial seeds that can be utilized by various state-of-the-art CLI mutators, including CarpetFuzz, ZigZagFuzz, and SelectFuzz.

\noindent{\bf{Extracting command-line options.}} Different mutators require seeds in specific formats. For example, CarpetFuzz~\cite{wang2023carpetfuzz} expects a single-line format: \texttt{\{command\} [options] -i @@}, where \texttt{@@} is replaced by the fuzzer with mutated input files. Converting complex shell scripts into this format requires semantic understanding of the script structure. 
\textcolor{black}{For example, Figure~\ref{fig:generated-script} shows a shell script generated by \sys.}
\sys uses the LLM to extract the essential command that reaches the target function and format it as a fuzzer seed, \texttt{ffmpeg -re -c:v copy -f rtp rtp://127.0.0.1:5030 -i @@}.
The LLM identifies the relevant program invocation, removes auxiliary commands (e.g., \texttt{timeout}), and reformats the command-line arguments into the fuzzer's expected format. This semantic extraction enables \sys to handle diverse shell script structures automatically.

\noindent{\bf{Collecting input files.}}
To obtain the input files referenced in each test, \sys executes all generated shell scripts (\texttt{run\_test1.sh}, \texttt{run\_test2.sh}, etc.) once and collects their outputs. For the example above, executing the script produces \texttt{input.mp4}, which becomes the corresponding seed file.

\section{Implementation and Experimental Setup}
This section describes the implementation and experimental setup.

\begin{table}[t]
\centering \setlength{\tabcolsep}{2pt} 
\caption{POWER dataset for baseline comparison.}
\label{target}
\footnotesize
\begin{tabular}{l|rrrr}
\toprule
\textbf{Program} & \textbf{LoC} & \textbf{\# Files} & \textbf{\# Functions}  & \textbf{Package name}\\
\midrule
avconv~\cite{libav2021}  & 680,276 & 1,193 & 11,074 & libav-git-c464278\\
bison~\cite{bison2021} & 530,528 & 133 & 1,497 & bison-3.7.6\\
cflow~\cite{cflow2021} & 80,174 & 28 & 400 & cflow-1.6\\
cjpeg~\cite{libjpegturbo2021}  & 112,689 & 39 & 312 & libjpeg-turbo-2.1.0\\
dwarfdump~\cite{libdwarf2021} & 183,051 & 109 & 1,503 & libdwarf-20210528\\
ffmpeg~\cite{ffmpeg2021} & 1,611,669 & 2,040 & 20,448 & ffmpeg-N-103440\\
gm~\cite{graphicsmagick2021} & 587,458 & 161 & 1,727 & GraphicsMagick-1.3.36\\
gs~\cite{ghostscript2021} & 3,527,477 & 1,497 & 21,673 & ghostpdl-9.54.0\\
mpg123~\cite{mpg123_2021}  & 113,461 & 70 & 733 & mpg123-1.28.2\\
nasm~\cite{nasm2021} & 158,205 & 79 & 832 & nasm-2.15.05\\
objdump~\cite{binutils2021} & 4,628,717 & 136 & 2,977 & binutils-2.36.1\\
pspp~\cite{pspp2021}  & 732,585 & 435 & 4,765 & pspp-1.4.1\\
readelf~\cite{binutils2021} & 4,628,717 & 130 & 1,289 & binutils-2.36.1\\
tiff2pdf~\cite{libtiff2021} & 173,153 & 34 & 846 & tiff-4.3.0\\
tiff2ps~\cite{libtiff2021} & 173,153 & 38 & 867 & tiff-4.3.0\\
vim~\cite{vim2021} & 918,746 & 135 & 6,864 & vim-8.2.3113\\
xmllint~\cite{libxml2_2021} & 549,947 & 45 & 3,070 & libxml2-2.9.12\\
xmlwf~\cite{expat2021} & 67,906 & 10 & 417 & expat-2.4.1\\
yara~\cite{yara2021} & 62,743 & 64 & 603 & yara-4.1.1\\
\bottomrule
\end{tabular}
\end{table}

\vspace{0.05in}
\noindent{\bf Implementation.} \hspace{0.05in}
Our implementation contains 12,872 Lines of Python Code. The default LLM is based on Claude Sonnet 4.0 (claude-sonnet-4-20250514)~\cite{3_7_sonnet}.
GPT-4.1 (gpt-4.1)~\cite{gpt} is also employed for the comparison. 
\textcolor{black}{\sys uses Clang's LibTooling API~\cite{libtooling} to construct call graphs by traversing the Abstract Syntax Tree (AST) of the target C source code.}
It also uses the NetworkX (nx)~\cite{NetworkX} library for graph analysis and centrality calculations. 

\vspace{0.05in}
\noindent{\bf Setup.} \hspace{0.05in}
We conducted our experiments on Microsoft Azure virtual machines. Seed generation was performed on a Standard\_L8as\_v3 instance with 8 vCPUs, 64 GB RAM, and 512 GB storage running Ubuntu 22.04 LTS. Fuzzing campaigns were executed on a Standard\_F32s\_v2 instance with 32 vCPUs, 64 GB RAM, and 512 GB storage running Ubuntu 20.04 LTS.
 We set the maximum number of iterative refinements per function ($N_{trial}$) to 2, the number of target functions ($N_{target}$) to 10, \textcolor{black}{and the maximum number of branch refinement attempts ($N_{branch}$) to 2}.
To reduce probabilistic behavior, the temperature~\cite{temperature} is set to 0, and \texttt{max\_tokens} parameter~\cite{temperature}, which indicates the maximum output tokens, is set to 4,096.
All target programs were compiled with AddressSanitizer~\cite{serebryany2012addresssanitizer} enabled to detect memory safety violations including buffer overflows, use-after-free, and NULL pointer dereferences.

\vspace{0.05in}
\noindent{\bf Dataset.} \hspace{0.05in}
We selected 43 C programs in total for our evaluation, consisting of three groups:

\noindent\textbf{(i) POWER Dataset (20 programs).}
We first included the 20 programs from the POWER~\cite{lee2022power} dataset, which has been widely used for command-line fuzzing evaluation in prior work including CarpetFuzz~\cite{wang2023carpetfuzz}, ProphetFuzz~\cite{wang2024prophetfuzz}, and ZigZagFuzz~\cite{lee2025zigzagfuzz}. These programs represent diverse applications ranging from parsers to multimedia tools and network utilities, with codebases between 62,743 and over 1,000,000 lines of code. 

\noindent\textbf{(ii) CarpetFuzz Dataset (13 programs).}
To enable comprehensive comparison with CarpetFuzz, we included the 13 additional programs that CarpetFuzz and ProphetFuzz evaluated.

\noindent\textbf{(iii) Extended Dataset (10 programs).}
To evaluate \sys's ability to discover zero-day vulnerabilities in actively maintained software, we selected 10 additional open-source projects based on two criteria: (1) high community adoption (GitHub stars $> 1,000$) and (2) recent developer activity (commits within the past 6 months). 
This extended evaluation is enabled by \sys's key advantage: unlike CarpetFuzz, which relies solely on publicly available initial seeds and thus cannot evaluate arbitrary programs, \sys generates seeds automatically for any target program.

For groups (i) and (ii), we used the same older versions as prior work to enable fair performance comparison. However, to discover zero-day vulnerabilities, we also evaluated these 33 programs at their latest versions. Combined with the 10 programs in group (iii), we performed zero-day vulnerability analysis on all 43 programs at their latest versions. 
Table~\ref{target} summarizes the 20 POWER dataset programs, while the remaining 23 programs are detailed in the artifact.
\begin{table}[t]
\centering
\caption{Comparison of initial seed generation methods and mutation strategies.}
\label{tab:fuzzer_comparison}
\begin{tabular}{lll}
\toprule
\textbf{Approach} & \textbf{Initial seed generation} & \textbf{Mutation} \\
\midrule
CarpetFuzz~\cite{wang2023carpetfuzz} & CarpetFuzz & CarpetFuzz \\
ProphetFuzz~\cite{wang2024prophetfuzz} & ProphetFuzz & CarpetFuzz \\
\sys-CF & \sys & CarpetFuzz \\
\midrule
ZigZagFuzz~\cite{lee2025zigzagfuzz} & (Default) dictionary & ZigZagFuzz \\
\sys-ZZ & \sys & ZigZagFuzz \\
\midrule
SelectFuzz~\cite{luo2023selectfuzz} & No initial seeds & SelectFuzz \\
\sys-SF & \sys & SelectFuzz \\
\bottomrule
\end{tabular}
\end{table}


\vspace{0.05in}
\noindent{\bf Baselines.} \hspace{0.05in}
In the evaluation, we conduct comprehensive experiments by integrating it with four state-of-the-art fuzzing frameworks.
\begin{itemize}
\item CarpetFuzz~\cite{wang2023carpetfuzz}: A state-of-the-art non-LLM 
approach that provides both initial seed generation and mutation capabilities. It leverages natural language processing to analyze command-line option inter-relationships from documentation for seed generation, and employs grammar-based mutations for fuzzing.
\item ProphetFuzz~\cite{wang2024prophetfuzz}: A state-of-the-art LLM-based approach for initial seed generation that leverages LLM's knowledge on high-risk option combinations.
\item ZigZagFuzz~\cite{lee2025zigzagfuzz}: A state-of-the-art CLI fuzzing mutator that variates both option configurations and input files.
\item SelectFuzz~\cite{luo2023selectfuzz}: A state-of-the-art directed fuzzing approach that selectively guides exploration toward specific target locations by identifying and prioritizing relevant code paths. \textcolor{black}{We selected SelectFuzz over AFLGo~\cite{bohme2017directed} as our directed fuzzing baseline because SelectFuzz is a more recent approach that extends distance-based guidance with precondition-guided path selection.}
\end{itemize}

Table~\ref{tab:fuzzer_comparison} summarizes our evaluation setup. We integrate \sys's seed generation with each mutation method to evaluate its effectiveness across different mutation strategies. 
Each fuzzing setup was run for 24 hours with \textcolor{black}{20} independent trials per program, and we measure the average coverage across these runs.
\textcolor{black}{To assess statistical significance, we performed Mann-Whitney U tests with Vargha-Delaney A12 effect sizes.}
\textcolor{black}{Each fuzzer run is executed on a dedicated Virtual Machine (VM) instance to eliminate resource contention. We report averages over independent trials to mitigate the effect of run-to-run variance. \sys's seed generation takes 28 minutes per program on average, and this seed generation time has already been included in the 24-hour budget.}
For CarpetFuzz mutator, we compare \sys-CF against two baseline seed generation approaches: the native CarpetFuzz method and ProphetFuzz. ProphetFuzz builds upon CarpetFuzz's mutator as its default base and provides a new initial seed generation method.
For ZigZagFuzz mutator, we compare \sys-ZZ against a standard dictionary-based approach (ZigZagFuzz). For SelectFuzz, which originally operates without initial seeds, we evaluate whether providing seeds generated by \sys (i.e., \sys-SF) can improve performance over the vanilla SelectFuzz.

To enable CLI fuzzing across all approaches, we apply the following modifications. For ZigZagFuzz, while it can leverage built-in initial seed options of target programs, these options are program-specific and not universally applicable. We therefore prepare a default dictionary by extracting options from program help screens.
For input files, we use the same default seed files from the fuzzing framework (AFL++'s basic seed corpus).
For SelectFuzz, which by default lacks an interface for mutating option strings, we integrate AFL++'s argv fuzzing interface~\cite{aflplusplus_argv} into its fuzzer component.


\begin{table}[t]
\centering
\caption{Target selection strategies evaluated in our experiments.}
\label{tab:centrality_metrics}
\footnotesize
\begin{tabular}{llp{4cm}}
\toprule
\textbf{Strategy} & \textbf{Metric} & \textbf{Description} \\
\midrule
CLOSE & Closeness centrality & Measures average distance from a function to all other functions in the call graph \\
BET & Betweenness centrality & Quantifies how often a function appears on shortest paths between other function pairs \\
DEG & Degree centrality & Counts the number of direct caller and callee connections \\
PAGE & PageRank & Assesses function influence based on the importance of its callers \\
random & Random & Randomly selects functions without considering graph structure\\
\bottomrule
\end{tabular}
\end{table}

\vspace{0.05in}
\noindent{\bf Preliminary Experiment.}
To analyze function call graph properties with respect to coverage performance and establish an effective function prioritization strategy, we conducted a preliminary experiment on the 20 POWER dataset programs. For each program, we selected 10 target functions per strategy using five different approaches: four commonly used centrality metrics (closeness centrality, betweenness centrality, degree centrality, and PageRank) and random selection (summarized in Table~\ref{tab:centrality_metrics}). We then instructed the LLM to generate test cases to reach those functions.
The results revealed that different centrality metrics excel under distinct structural conditions. Based on these findings, we developed automated decision rules for \sys's target selection mechanism that select the most effective strategy among the five approaches for each program based on its structural characteristics. Section~\ref{sec:rq4-2} presents the detailed analysis and validation of this mechanism's contribution to coverage improvement.

\vspace{0.05in}
\noindent{\bf Evaluation Metrics.} \hspace{0.05in}
We evaluate \sys and the baselines based on the following metrics:
\begin{itemize}
\item Vulnerability detection: Number of vulnerabilities that are detected by \sys
\item Edge coverage: \textcolor{black}{Number of unique edge transitions between basic blocks observed during fuzzing, as recorded by AFL++'s compile-time instrumentation via a shared bitmap.}
\end{itemize}
\textcolor{black}{Regarding edge coverages, all approaches in our evaluation use the same instrumentation, ensuring a fair comparison. We note that AFL++'s bitmap-based tracking may experience hash collisions for programs with very large CFGs, but this affects all compared approaches equally and does not impact relative comparisons.}


\section{Evaluation}\label{sec:evaluation}
We address the following research questions:

\begin{itemize}
    \item \textbf{RQ1} [Zero-day Vulnerabilities] How many zero-day vulnerabilities can \sys discover? 
    \item \textbf{RQ2} [Vulnerability Detection] How does \sys compare to baselines in terms of vulnerability detection?
    \item \textbf{RQ3} [Code Coverage] How does \sys compare with baselines in terms of code coverage? 
    \item \textbf{RQ4} [Ablation study] How does each strategy of \sys contribute to the coverage metrics?   
    \item \textbf{RQ5} [Cost] What are the costs of \sys?
\end{itemize}

\begin{table}[t]
\centering
\caption{[RQ1] Zero-day vulnerabilities discovered by \sys.}
\label{tab:bug_types}
\begin{tabular}{lrrr}
\toprule
\textbf{Vulnerability type} & \textbf{Count} & \textbf{Confirmed} & \textbf{Fixed} \\
\midrule
Buffer overflow & 16 & 14 & 11 \\
NULL pointer dereference & 12 & 10 & 8 \\
Memory leak & 8 & 4 & 3 \\
Out-of-bounds read & 4 & 4 & 2 \\
Use-after-free & 2 & 2 & 2 \\
Invalid pointer access & 2 & 2 & 2 \\
Double-free & 1 & 0 & 0 \\
Division by zero & 1 & 1 & 1 \\
Integer overflow & 1 & 1 & 1 \\
Others\textsuperscript{\dag} & 4 & 3 & 3 \\
\midrule
\textbf{Total} & 51 & 41 & 33 \\
\bottomrule
\multicolumn{4}{p{0.9\linewidth}}{\footnotesize \textsuperscript{\dag} Heap address leak (1), memory exhaustion (1), infinite loop (1), reachable assertion (1)} \\
\end{tabular}
\end{table}

\subsection{RQ1: Zero-day Vulnerabilities}~\label{rq1}

To evaluate the effectiveness of discovering previously unknown vulnerabilities, we applied \sys to all 43 programs at their latest versions.

\vspace{0.05in}
\noindent{\bf Vulnerability Discovery.} \hspace{0.05in}
\sys identified 51 previously unknown vulnerabilities across 24 programs. 
Table~\ref{tab:bug_types} summarizes the distribution of vulnerability types.
The discovered vulnerabilities include 16 buffer overflows, 12 NULL pointer dereferences, 8 memory leaks, 4 out-of-bounds reads, 2 use-after-free cases, and several other logical flaws.
Of these, 41 have been confirmed by maintainers, with 33 already fixed in subsequent releases.
\textcolor{black}{Five} vulnerabilities have been assigned CVEs.
The complete list of vulnerabilities with detailed information is provided in the artifact.



\begin{table*}[t]
\centering
\small
\caption{[RQ2] Vulnerability detection performance comparison across 33 baseline programs (24-hour runs). Unique vulnerabilities refer to vulnerabilities not detected by any other method in this comparison.}
\label{tab:comparison_results}
\begin{tabularx}{\textwidth}{lRRRRRRR}
\toprule
& \multicolumn{3}{c@{\hspace{12pt}}}{\textbf{CF mutation}} & \multicolumn{2}{c@{\hspace{12pt}}}{\textbf{ZZ mutation}} & \multicolumn{2}{c}{\textbf{SF mutation}}\\
\cmidrule(lr){2-4} \cmidrule(lr){5-6} \cmidrule(lr){7-8}
\textbf{} & \textbf{CarpetFuzz} & \textbf{ProphetFuzz} & \textbf{\sys-CF} & \textbf{ZigZagFuzz} & \textbf{\sys-ZZ} & \textbf{SelectFuzz} & \textbf{\sys-SF}\\
\midrule
Total vulnerabilities & \textcolor{black}{26} & 7 & \textbf{33} & 18 & \textbf{25} & 0 & \textbf{3} \\
Unique vulnerabilities & 18 & 4 & 27 & 4 & 6 & 0 & 1 \\
\midrule
Memory corruption & 16 & 0 & 11 & 4 & 7 & 0 & 0 \\
NULL pointer dereferences & 10 & 6 & 18 & 11 & 10 & 0 & 1 \\
Invalid operations & 0 & 0 & 2 & 0 & 4 & 0 & 0 \\
Resource management & 0 & 1 & 1 & 2 & 3 & 0 & 2 \\
Arithmetic/logic errors & 0 & 0 & 1 & 1 & 1 & 0 & 0 \\
\midrule
Call chain depth & 4.27 (±1.53) & 4.11 (±1.21) & \textbf{4.79 (±2.91)} & 2.22 (±1.03) & 3.39 (±1.89) & 0.00 & 2.00 (±1.41) \\
\bottomrule
\end{tabularx}
\vspace{-2mm}
\end{table*}


\vspace{0.05in}
\noindent{\bf Representative Case Studies.} \hspace{0.05in}
To illustrate the types of vulnerabilities discovered by \sys, we present two representative case studies: a buffer underflow in ImageMagick and a buffer overflow in YARA.
We verified that state-of-the-art fuzzing approaches could not discover these vulnerabilities within the same time budget.

\noindent{\bf Case Study 1: Buffer Underflow in ImageMagick.} \hspace{0.05in}
\sys identified a buffer underflow vulnerability \textcolor{black}{(CVE-2025-53101)} in ImageMagick, specifically within the filename template processing logic of the \texttt{InterpretImageFilename()} function in \texttt{image.c}.
The vulnerability stems from an incorrect assumption in the offset calculation mechanism. 
When format specifiers beginning with \texttt{\%} are processed consecutively, the cumulative offset exceeds the template position difference, causing the buffer address calculation to underflow to a location preceding the buffer's start.
Although not explicitly documented in the ImageMagick command documentation, such format specifiers are semantically necessary arguments for the filename templating functionality. 
\sys included the \texttt{\%} format specifier in its seed (e.g., \texttt{-format "\%w x \%h"}), enabling the discovery of this vulnerability that other approaches missed.


\noindent{\bf Case Study 2: Buffer Overflow in YARA.} \hspace{0.05in}
\sys identified a heap buffer overflow vulnerability \ts{(CWE-122)} in YARA, specifically within the \texttt{yr\_arena\_load\_stream()} function in \texttt{libyara/arena.c}.
The vulnerability arises from integer underflow when validating buffer boundaries: small buffer sizes cause arithmetic underflow that bypasses size checks, enabling out-of-bounds memory access.

A critical challenge in reaching this function was that it only processes compiled YARA rule files (\texttt{.yarac}), which must be generated using the \texttt{yarac} compiler from source \texttt{.yar} files. Traditional seed generation approaches cannot orchestrate auxiliary tools like \texttt{yarac} to create the required input format. \sys \ts{targeted the \texttt{yr\_arena\_load\_stream()} function,} analyzed the call paths and identified this dependency, then generated a complete workflow: creating \texttt{.yar} rule files, compiling them with \texttt{yarac} to produce \texttt{.yarac} files, and invoking \texttt{yara -C} to load the compiled rules. 
\sys selected \ts{\texttt{yr\_arena\_load\_stream()}}{this function} as a target based on its high centrality in the call graph. Prior research~\cite{liu2020large} stated that high-centrality functions tend to contain more vulnerabilities, and this case validates that finding: the target function indeed contained a vulnerability.

\findingbox{\sys discovered 51 previously unknown vulnerabilities across 24 programs, with 41 confirmed and 33 fixed by developers, including \textcolor{black}{five} assigned CVE identifiers.}




\begin{table*}[t]
\centering
\caption{\textcolor{black}{[RQ3] Edge coverage comparison across 20 POWER programs after 24 hours of fuzzing. Bold values indicates the best-performing baseline (CarpetFuzz or ProphetFuzz for CF-based; ZigZagFuzz for ZZ-based; SelectFuzz for SF-based). ``--'' indicates that the baseline could not generate initial seeds due to insufficient information in man pages.}}
\label{tab:edge_coverage}
\footnotesize
\begin{tabular}{l@{\hspace{8pt}}rrr@{\hspace{12pt}}rr@{\hspace{12pt}}rr}
\toprule
& \multicolumn{3}{c@{\hspace{12pt}}}{\textbf{CF mutation}} & \multicolumn{2}{c@{\hspace{12pt}}}{\textbf{ZZ mutation}} & \multicolumn{2}{c}{\textbf{SF mutation}}\\
\cmidrule(lr){2-4} \cmidrule(lr){5-6} \cmidrule(lr){7-8}
\textbf{Program} & \textbf{CarpetFuzz} & \textbf{ProphetFuzz} & \textbf{\sys-CF} & \textbf{ZigZagFuzz} & \textbf{\sys-ZZ} & \textbf{SelectFuzz} & \textbf{\sys-SF} \\
\midrule
avconv & 10,832$\pm$3,861 & 22,515$\pm$4,494 & \textbf{30,074$\pm$4,549} & 15,776$\pm$4,452 & \textbf{25,493$\pm$3,928} & 0$\pm$0 & \textbf{460$\pm$163} \\
bison & 5,792$\pm$206 & 4,720$\pm$783 & \textbf{6,651$\pm$244} & 7,217$\pm$745 & \textbf{7,301$\pm$621} & 2,310$\pm$371 & \textbf{3,428$\pm$135} \\
cflow & 1,647$\pm$11 & 1,312$\pm$31 & \textbf{1,694$\pm$7} & 2,208$\pm$91 & \textbf{2,263$\pm$38} & 563$\pm$39 & \textbf{702$\pm$16} \\
cjpeg & 737$\pm$253 & 1,133$\pm$81 & \textbf{1,451$\pm$178} & 4,556$\pm$1,755 & \textbf{5,732$\pm$269} & 57$\pm$75 & \textbf{339$\pm$43} \\
dwarfdump & 6,251$\pm$348 & 5,267$\pm$416 & \textbf{7,310$\pm$639} & 8,194$\pm$837 & \textbf{9,451$\pm$645} & 51$\pm$10 & \textbf{128$\pm$258} \\
ffmpeg & 20,856$\pm$5,889 & 27,092$\pm$6,855 & \textbf{43,572$\pm$3,244} & 26,846$\pm$9,723 & \textbf{34,734$\pm$7,474} & 10$\pm$0 & \textbf{140$\pm$112} \\
gm & 5,945$\pm$1,116 & 5,924$\pm$1,045 & \textbf{9,809$\pm$1,350} & 5,158$\pm$1,468 & \textbf{7,257$\pm$2,263} & 569$\pm$153 & \textbf{3,734$\pm$420} \\
gs & 19,327$\pm$764 & 16,652$\pm$889 & \textbf{21,867$\pm$1,073} & 16,684$\pm$1,697 & \textbf{16,839$\pm$530} & 441$\pm$59 & \textbf{1,573$\pm$160} \\
jasper & 1,867$\pm$377 & -- & \textbf{2,269$\pm$19} & 3,076$\pm$412 & \textbf{3,990$\pm$510} & 12$\pm$0 & \textbf{576$\pm$249} \\
mpg123 & 2,834$\pm$258 & 2,432$\pm$245 & \textbf{3,966$\pm$276} & 3,840$\pm$284 & \textbf{3,991$\pm$479} & 21$\pm$36 & \textbf{31$\pm$81} \\
nasm & 6,945$\pm$287 & 5,317$\pm$253 & \textbf{7,607$\pm$88} & \textbf{8,648$\pm$539} & 5,983$\pm$670 & 527$\pm$140 & \textbf{1,250$\pm$122} \\
objdump & 6,954$\pm$1,308 & 8,673$\pm$1,515 & \textbf{11,596$\pm$1,777} & 6,549$\pm$1,205 & \textbf{11,135$\pm$1,189} & 84$\pm$11 & \textbf{2,212$\pm$295} \\
pspp & 4,507$\pm$2,160 & -- & \textbf{10,752$\pm$1,013} & 3,303$\pm$709 & \textbf{9,856$\pm$739} & 1,379$\pm$167 & \textbf{3,301$\pm$82} \\
readelf & 6,875$\pm$1,093 & 6,219$\pm$1,069 & \textbf{9,342$\pm$1,717} & 6,227$\pm$2,413 & \textbf{9,527$\pm$2,149} & 1$\pm$0 & \textbf{547$\pm$13} \\
tiff2pdf & 3,069$\pm$470 & 3,897$\pm$360 & \textbf{4,650$\pm$543} & 3,067$\pm$564 & \textbf{3,962$\pm$863} & 208$\pm$356 & \textbf{868$\pm$60} \\
tiff2ps & 2,164$\pm$439 & 2,727$\pm$269 & \textbf{3,184$\pm$304} & 2,707$\pm$348 & \textbf{3,251$\pm$465} & 81$\pm$102 & \textbf{171$\pm$22} \\
vim & 22,623$\pm$3,261 & 15,345$\pm$7,485 & \textbf{25,319$\pm$10,414} & 37,098$\pm$7,514 & \textbf{38,812$\pm$4,995} & \textbf{2,039$\pm$75} & 1,632$\pm$218 \\
xmllint & 8,632$\pm$628 & 6,681$\pm$651 & \textbf{9,885$\pm$858} & 14,065$\pm$2,196 & \textbf{15,984$\pm$2,749} & 424$\pm$263 & \textbf{2,781$\pm$1,457} \\
xmlwf & 3,637$\pm$203 & 3,348$\pm$222 & \textbf{3,832$\pm$389} & 3,602$\pm$691 & \textbf{4,030$\pm$439} & 164$\pm$27 & \textbf{173$\pm$14} \\
yara & 2,410$\pm$159 & 1,734$\pm$266 & \textbf{2,786$\pm$69} & 2,635$\pm$239 & \textbf{3,256$\pm$249} & 234$\pm$48 & \textbf{801$\pm$43} \\
\bottomrule
\end{tabular}
\vspace{-2mm}
\end{table*}

\subsection{RQ2: Vulnerability Detection \ts{Performance Comparison}}



To evaluate \sys's vulnerability detection capability against prior work, we conducted a comparative evaluation using the same program versions from the POWER and CarpetFuzz datasets and configurations as baseline fuzzers.

\autoref{tab:comparison_results} presents the vulnerability detection results across all baselines.
\sys consistently outperforms baselines in total vulnerabilities discovered: 33 vs \textcolor{black}{26} (CarpetFuzz), 33 vs 7 (ProphetFuzz), 25 vs 18 (ZigZagFuzz), and 3 vs 0 (SelectFuzz). 
\sys discovers substantially more unique vulnerabilities, those not found by any other method in this comparison: 27 vs 18 (CarpetFuzz), 27 vs 4 (ProphetFuzz), 6 vs 4 (ZigZagFuzz), and 1 vs 0 (SelectFuzz). This demonstrates that \sys explores distinct code regions that baseline approaches miss, rather than merely finding overlapping vulnerabilities. 

\sys's semantic exploration approach involves trade-offs with systematic numerical enumeration. 
For example, CarpetFuzz excels at discovering edge cases in specific numerical ranges through exhaustive exploration. It finds tiffcrop vulnerabilities by systematically testing coordinate combinations (e.g., \texttt{-z 1,1,2048,2048:1,2049,2048,4097}) and rotation angles (\texttt{-R 90}, \texttt{-R 270}). 
While \sys fails to generate these precise numerical patterns, as illustrated in the case study in the previous section, it compensates by producing diverse semantically plausible inputs, enabling discovery of vulnerabilities in complex logic paths that numerical enumeration misses.


Beyond vulnerability quantity, \sys reaches deeper code locations. Call chain depth in ~\autoref{tab:comparison_results} measures the average number of function calls from the program entry point to the vulnerability location. \sys-CF achieves an average depth of 4.79 (±2.91) compared to 4.27 (±1.53) for CarpetFuzz, with vulnerabilities discovered across a wider depth range (1-10 vs 1-7).
The large standard deviation for \sys-CF reflects its ability to discover vulnerabilities across a wider range of call depths (1 to 10) than CarpetFuzz (1 to 7).
Similarly, \sys-ZZ reaches depth 3.39 compared to 2.22 for ZigZagFuzz, consistently demonstrating \sys's ability to explore deeply nested functions.

These results also reveal that the CarpetFuzz mutator reaches deeper code locations than the ZigZagFuzz mutator. This is because CarpetFuzz prepares multiple option seeds but does not mutate the strings themselves, thereby preserving the inter-relationships between options. Since \sys generates option strings based on contextual understanding, its seeds are particularly effective when paired with mutators that maintain these option relationships, as evidenced by the superior performance of \sys-CF over \sys-ZZ.

\findingbox{\sys discovers 1.4-4.7× more total vulnerabilities and 1.5-6.8× more unique vulnerabilities than baselines, reaching deeper code locations through path-guided exploration of distinct code regions.}

\begin{table}[t]
\centering
\caption{\textcolor{black}{Edge coverage comparison between PILOT and ProphetFuzz both using gpt-4 with CarpetFuzz mutation (24-hour runs).}}
\label{tab:gpt4_comparison}
\textcolor{black}{
\begin{tabular}{lrr}
\toprule
Program & ProphetFuzz & PILOT-GPT4 \\
\midrule
avconv & 22,515$\pm$4,494 & \textbf{33,521$\pm$2,052} \\
bison & 4,720$\pm$783 & \textbf{5,387$\pm$123} \\
cflow & 1,312$\pm$31 & \textbf{1,589$\pm$5} \\
cjpeg & 1,133$\pm$81 & \textbf{1,293$\pm$21} \\
dwarfdump & 5,267$\pm$416 & \textbf{6,542$\pm$86} \\
ffmpeg & 27,092$\pm$6,855 & \textbf{45,055$\pm$1,419} \\
gm & 5,924$\pm$1,045 & \textbf{12,565$\pm$465} \\
gs & 16,652$\pm$889 & \textbf{16,824$\pm$1,715} \\
mpg123 & 2,432$\pm$245 & \textbf{3,652$\pm$59} \\
nasm & \textbf{5,317$\pm$253} & 4,930$\pm$194 \\
objdump & 8,673$\pm$1,515 & \textbf{11,598$\pm$554} \\
readelf & 6,219$\pm$1,069 & \textbf{9,895$\pm$408} \\
tiff2pdf & 3,897$\pm$360 & \textbf{4,813$\pm$118} \\
tiff2ps & 2,727$\pm$269 & \textbf{3,714$\pm$56} \\
vim & 15,345$\pm$7,485 & \textbf{30,127$\pm$3,297} \\
xmllint & 6,681$\pm$651 & \textbf{10,783$\pm$279} \\
xmlwf & 3,348$\pm$222 & \textbf{4,001$\pm$27} \\
yara & 1,734$\pm$266 & \textbf{2,524$\pm$30} \\
\bottomrule
\end{tabular}
}
\end{table}

\subsection{RQ3: Coverage Improvement}
\noindent{\textbf{Fuzzing Coverage.}}
\autoref{tab:edge_coverage} presents edge coverage results for a representative subset of 20 POWER baseline programs.
\textcolor{black}{
Across the full 33-program benchmark suite, \sys demonstrates superior coverage, achieving average improvements of 20\% over CarpetFuzz, 28\% over ProphetFuzz, 15\% over ZigZagFuzz, and 66\% over SelectFuzz. 
Mann-Whitney U tests confirm that these improvements are statistically significant: 90 \% of program-baseline pairs show significant improvement ($p < 0.05$), with a median Vargha-Delaney A12 effect size of 0.97 (mean: 0.88), indicating a large effect. The detailed statistical results are presented in Table~\ref{tab:stat-all} in Appendix~\ref{app:stats}.}

The performance gap is particularly large for SelectFuzz. 
SelectFuzz uses distance metrics to guide mutations toward target locations, but struggles when initial seeds cannot reach the vicinity of target functions. 
CLI programs require specific option combinations to activate functionalities, creating all-or-nothing dependencies. 
Without initial seeds that satisfy these dependencies, SelectFuzz's distance-guided approach cannot compute meaningful guidance metrics, significantly limiting its effectiveness.

\textcolor{black}{To ensure a fair comparison controlling for LLM capability, we also evaluated both \sys and ProphetFuzz using gpt-4 (ProphetFuzz's default LLM) with CarpetFuzz mutation. 
As shown in Table~\ref{tab:gpt4_comparison}, on the 18 programs where both ProphetFuzz and PILOT-GPT-4 have sufficient data, \sys-GPT-4 achieves 35\% higher coverage than ProphetFuzz on average, demonstrating that \sys's improvement stems from its architecture rather than the choice of underlying LLM.}



\noindent{\textbf{Initial Seed Coverage.}}
To understand what drives these coverage improvements, we analyzed initial seed quality for each approach.
\autoref{tab:function_depth} compares the characteristics of initial seed sets by executing each seed once.

The \textit{minimal set} configuration represents the simplest possible invocation, containing only a single fixed option string, which is the typical approach used by many non-CLI-aware fuzzing frameworks~\cite{lee2025zigzagfuzz}. For input files, 
we use the AFL++ fuzzer's default seed files for this minimal set configuration.
Comparing against the minimal set (261 functions, 1,446 branches), we observe that CLI-aware approaches achieve substantially higher coverage. CarpetFuzz achieves 1.59× more functions, ProphetFuzz achieves 1.41×, and \sys achieves 2.33×, demonstrating that systematic CLI option exploration is essential for reaching diverse program functionality.

Then, \sys substantially outperforms all existing methods. Compared to CarpetFuzz, \sys achieves 1.47× more function coverage, and 1.64× more branch coverage.
Dictionary-based fuzzing performs worst with only 242 functions due to its limited predefined dictionary, while ProphetFuzz's LLM-guided approach achieves 369 functions but falls short of \sys's coverage-guided approach.

To isolate individual contributions of CLI option generation and input file generation, we evaluate two ablated versions: fixed option (using \sys's input files but single (minimal) option) achieves 340 functions, while fixed files (using \sys's CLI generation but minimal seed files) achieves 509 functions. This reveals that CLI option generation contributes 1.96× improvement while input file generation contributes 1.79×. The combination produces synergistic benefits, with \sys achieving the highest coverage and deepest call depth (4.77) across all metrics.

Table~\ref{tab:option_discovery_summary} provides a comprehensive view of seed generation across all 33 evaluated programs. \sys discovers more CLI options than existing approaches (1,026 vs 629 for CarpetFuzz).
Importantly, \sys discovers 553 unique options not found by other approaches, demonstrating its superior ability to explore diverse CLI interfaces. Additionally, \sys generates significantly more input files, producing an average of 204.3 files per program compared to 12.9 for CarpetFuzz and 44.1 for ProphetFuzz, which enables more comprehensive testing of file-dependent functionalities.


\begin{table}[t]
\centering
\caption{Branch coverage, function coverage, and call depth statistics from execution of initial seeds.}
\label{tab:function_depth}
\small
\begin{tabular}{lcccc}
\toprule
\textbf{Approach} & \textbf{Branches} & \textbf{Functions}  & \textbf{Avg depth} \\
\midrule
Minimal set        & 1,446 ± 2,290 & 261 ± 435 & 4.21 \\
\midrule
Dictionary            & 1,365 ± 2,165 & 242 ± 415 & 3.82 \\
CarpetFuzz          & 2,740 ± 3,499  & 414 ± 559 &  4.27 \\
ProphetFuzz        & 2,506 ± 2,723  & 369 ± 420 & 4.68 \\
\sys          & \textbf{4,487 ± 5,087}  & \textbf{609 ± 794} &  \textbf{4.77} \\
\midrule
Fixed option    & 2,045 ± 2,076  & 340 ± 454  &  4.47 \\
Fixed files      & 3,592 ± 5,352  & 509 ± 809  & 4.31 \\
\bottomrule
\end{tabular}
\end{table}

\begin{table}[t]
\centering
\caption{Initial seed set characteristics: generated input files and discovered options.}
\label{tab:option_discovery_summary}
\small
\begin{tabular}{l@{\hspace{3pt}}c@{\hspace{3pt}}c@{\hspace{3pt}}c}
\toprule
\textbf{Approach} & \textbf{Files} & \textbf{Options}  & \textbf{Unique options} \\
\midrule
Dictionary   & -- & 771 & 329 \\
CarpetFuzz  & 12.9 & 629  & 176 \\
ProphetFuzz  & 44.1 & 595 & 187 \\
\sys         & \textbf{204.3} & \textbf{1,026}  & \textbf{553} \\
\bottomrule
\end{tabular}
\end{table}

To illustrate \sys's semantic advantage concretely, we categorize \sys's unique options into three types:

\begin{table*}[t]
\centering
\caption{[RQ4-1] Function and branch coverage for different configurations.}
\footnotesize
\begin{tabular}{l@{\hspace{8pt}}rr@{\hspace{10pt}}rr@{\hspace{10pt}}rr@{\hspace{10pt}}rr@{\hspace{10pt}}rr}
\toprule
& \multicolumn{2}{c@{\hspace{10pt}}}{\textbf{W/o path-guided}} & \multicolumn{2}{c@{\hspace{10pt}}}{\textbf{W/o refinement}} & \multicolumn{2}{c@{\hspace{10pt}}}{\textbf{W/o config}} & \multicolumn{2}{c@{\hspace{10pt}}}{\textbf{\sys-GPT}} & \multicolumn{2}{c}{\textbf{\sys}} \\
\cmidrule(lr){2-3} \cmidrule(lr){4-5} \cmidrule(lr){6-7} \cmidrule(lr){8-9} \cmidrule(lr){10-11}
\textbf{Program} & \textbf{Function} & \textbf{Branch} & \textbf{Function} & \textbf{Branch} & \textbf{Function} & \textbf{Branch} & \textbf{Function} & \textbf{Branch} & \textbf{Function} & \textbf{Branch} \\
\midrule
cjpeg & 184 & 1,372 & 188 & 1,278 & 197 & 1,706 & 140 & 805 & \textbf{205} & \textbf{2,068} \\
dwarfdump & 628 & 4,164 & 615 & 3,806 & 808 & 5,178 & 543 & 3,058 & \textbf{918} & \textbf{6,439} \\
gm & 624 & 7,319 & 554 & 6,212 & \textbf{743} & 7,554 & 371 & 3,217 & 667 & \textbf{8,290} \\
jasper & 358 & 2,577 & 337 & 2,311 & 421 & 3,045 & 338 & 2,312 & \textbf{428} & \textbf{3,257} \\
nasm & 475 & 3,360 & 498 & 3,576 & 480 & 3,541 & 287 & 1,587 & \textbf{565} & \textbf{4,348} \\
objdump & 409 & 3,020 & 453 & 3,223 & 439 & 3,241 & 327 & 1,972 & \textbf{607} & \textbf{5,068} \\
pspp & 1,593 & 5,213 & 1,802 & 5,746 & 1,755 & 5,616 & 1,112 & 3,094 & \textbf{1,904} & \textbf{6,311} \\
tiff2ps & 184 & 1,291 & 193 & 1,230 & 201 & 1,535 & 189 & 1,339 & \textbf{328} & \textbf{2,563} \\
xmlwf & 160 & 1,190 & 140 & 1,080 & 162 & 1,267 & 124 & 747 & \textbf{174} & \textbf{1,376} \\
cflow & 184 & 965 & 197 & 1,080 & 256 & 1,557 & 182 & 935 & \textbf{269} & \textbf{1,682} \\
\midrule
\textbf{Avg. change} & -19.8\% & -27.9\% & -17.9\% & -27.5\% & -11.4\% & -19.8\% & -36.6\% & -51.1\% & -- & -- \\
\midrule
\textbf{Func. reachability} & 51\% &  & 54\% &  & 59\% &  & 49\% &  & \textbf{64\%} &  \\
\bottomrule
\end{tabular}
\label{tab:ablation}
\end{table*}

\noindent{(i) Standard functional options:}
\sys extracts options directly from source code rather than relying on documentation, therefore it naturally discovers more options. 
For example, gs (Ghostscript) has extensive internal parameters for controlling rendering engine behavior, memory management, and color processing.
\sys discovers 214 options including these internal parameters such as \texttt{\mbox{-dGraphicsAlphaBits}} and \texttt{\mbox{-dNumRenderingThreads}}, while ProphetFuzz finds only 31 commonly documented options like \texttt{\mbox{-sDEVICE}} and \texttt{\mbox{-sOutputFile}}.

\noindent{(ii) Numeric boundary testing:}
\sys generates rich numeric variations targeting edge cases.
For example, in editcap, \sys generates timestamps including \texttt{-2147483648} (Unix epoch minimum), \texttt{-999999999}, and \texttt{-86400}.
This approach ensures comprehensive coverage of numeric edge cases.

\noindent{(iii) Invalid option generation:} \sys deliberately generates invalid options to test error handling robustness (e.g., \texttt{\mbox{-{}-nonexistent}} and \texttt{\mbox{-{}-unknown-param}}).
\sys guides LLMs to analyze the control flow logic for reaching error handling functions.


\findingbox{\textcolor{black}{\sys achieves substantial coverage improvements across the 33-program benchmark suite, with average gains of 20\% over CarpetFuzz, 28\% over ProphetFuzz, 15\% over ZigZagFuzz, and 66\% over SelectFuzz.}}

\subsection{RQ4: Ablation Study}~\label{ablation}
We conduct ablation studies to evaluate (1) the contribution of core \sys components and (2) the effectiveness of different target selection strategies in improving coverage.

\subsubsection{Core Component Contributions}
We conduct ablation studies on 10 programs randomly selected from our benchmark suite, with 5 target functions for each program.
We examined four configurations: without path-guided prompting, without iterative refinement, without native tool configuration, and with GPT-4.1 instead of Claude Sonnet 4.0.
\autoref{tab:ablation} shows the results.


\vspace{0.05in}
\noindent{\bf {W/o path-guided prompting.}} \hspace{0.05in}
This configuration disables the path candidate summary prompting.
On average, disabling the path-guided prompting resulted in a 19.8\% reduction in function coverage and a 27.9\% reduction in branch coverage across all programs.
Without path guidance, \sys must search the call sites of the target functions through the entire large codebase, significantly reducing its ability to generate effective test cases that cover the intended functionality.
This is further evidenced by the function reachability dropping from 64\% to 51\%, indicating that path-guided prompting plays a crucial role in helping the LLM successfully reach the target functions.



\vspace{0.05in}
\noindent{\bf {W/o iterative refinement.}}
This configuration disables \sys's coverage-driven feedback loop, generating test cases only once without refinement based on coverage results. Instead of 
iteratively querying the LLM with feedback about which functions along candidate paths were covered, this variant generates all seeds in a single attempt. 
On average, removing iterative refinement resulted in a 17.9\% reduction 
in function coverage and a 27.5\% reduction in branch coverage across all programs.

\vspace{0.05in}
\noindent{\bf {W/o native tool configuration.}}
\hspace{0.05in}
This configuration removes the instruction to use native command-line tools for input file generation from the LLM prompts.
This results in an average 11.4\% reduction in function coverage and a 19.8\% reduction in branch coverage, as the system cannot generate valid files for programs requiring specific file format inputs. 
Instead, it produces files with arbitrary bytes (e.g., \texttt{printf '\textbackslash x00\textbackslash x00\textbackslash x00\textbackslash x00' > input.bin}) that fail to satisfy format requirements.

\vspace{0.05in}
\noindent{\bf {Other LLM types beyond Claude.}}
\hspace{0.05in}
The default \sys with Claude achieves an average function coverage of 506.5 functions, while \sys-GPT achieves 361.3 functions. 
For branch coverage, \sys achieves an average of 4,232 branches compared to \sys-GPT's 1,907 branches.
This consistent advantage stems from Claude Sonnet 4.0's strengths in two key areas: (1) advanced coding capabilities that enable more accurate test case generation, and (2) superior performance in agentic workflows that require iterative reasoning and tool use.
\textcolor{black}{Specifically, under identical prompts, Claude actively reads source code (17.6 \% of API uses) to identify specific branch conditions before crafting targeted scripts, while GPT relies solely on the provided path context without additional code exploration (0 \% read usage), producing repetitive parameter variations.
}


\subsubsection{Effectiveness of Target Selection Strategies}~\label{sec:rq4-2}
To evaluate \sys's call graph-based target selection strategy, we analyzed the relationship between graph structural properties and coverage performance.

\noindent{\bf {Correlation analysis.}} \hspace{0.05in}
To identify when centrality-based strategies outperform random selection, we conduct correlation analysis between call graph structure and strategy. 
For each of the 20 POWER programs, we measure branch coverage achieved by each strategy at a fixed token budget (one million tokens, averaged over three trials). We then compute each strategy's \textit{advantage} over random selection:
$A(p,s) = C(p,s) - C(p,\texttt{random})$, where $C(p,s)$ is the coverage achieved by strategy $s$ on program $p$. A positive advantage indicates superior performance.

To understand what drives these advantages, we extract structural features from each program's call graph: basic metrics (nodes, edges, density), centrality distributions, and topological properties (clustering coefficient, diameter, strongly connected components). We then correlate these features with strategy advantages.

The correlation analysis reveals distinct structural preferences for each strategy (detailed in Appendix~\ref{app:correlations}). For example, CLOSE excels on programs with large diameter and long paths, where high-closeness functions bridge distant components. BET and DEG prefer concentrated PageRank distributions, succeeding when few functions dominate. PAGE favors fragmented graphs with small strongly connected components.


\noindent{\bf {Strategy results.}} \hspace{0.05in}
Based on the significant correlations, we generate rules for deciding strategies. For each strategy, we establish thresholds using median feature values where the strategy outperforms random selection, weighted by correlation strength (detailed in Appendix~\ref{app:decision_rules}). 
Our structural features successfully predict strategy effectiveness for all programs in the preliminary experiment. 
The average confidence score across recommendations was 0.80, indicating high prediction accuracy.
PAGE is most frequently chosen (40 \%), favoring fragmented call graphs where PageRank identifies important nodes in loosely connected structures. 
BET suits 35 \% of programs with concentrated PageRank distributions, while CLOSE is
\textcolor{black}{best suited for} 20 \% with large diameters where distance-based centrality matters most.

\findingbox{\sys outperforms alternative configurations in the ablation study. These results demonstrate that \sys's path-guided prompting, iterative feedback loop, native tool configuration, and adaptive target selection are essential components for achieving comprehensive program coverage.}

\subsection{RQ5: Cost \textcolor{black}{and Time} Analysis}
We analyzed the token usage and API costs across the POWER dataset (detailed results in Appendix~\ref{app:cost_details}). The results reveal clear patterns in LLM resource consumption for automated test generation.
On average, each program required 36 chat interactions with the LLM to complete the test generation process, consuming approximately 4.6 million input tokens and 90,000 output tokens per trial. 
The iterative nature of the exploration process, requiring 27-46 conversational exchanges per program, demonstrates how the LLM progressively builds understanding through multiple interactions with the codebase.
The input-to-output token ratio averages approximately 2\%, indicating that the LLM primarily operates in read mode to explore and gather necessary knowledge about the codebase, with relatively concise outputs for test case generation.

The average cost per program was \$15.16 (USD), with individual costs ranging from \$10.50 to \$20.05 depending on program complexity and the extent of code exploration required.
While this represents a substantial computational investment, it remains economically feasible for practical vulnerability discovery.

\textcolor{black}{As a comparison, ProphetFuzz incurs an average cost of \$8 per program. While PILOT's cost is higher, the coverage and vulnerability detection improvements indicate a favorable cost-effectiveness trade-off.}

\textcolor{black}{\sys's seed generation takes $28.0\pm3.9$ minutes per program on average, ranging 19.4–34.5 minutes across the 43 programs, which is relatively lightweight compared with fuzzing.}

\section{Related Work}\label{sec:relate}

\noindent{\bf {Command-line Fuzzing.}}
\textcolor{black}{AFLargv~\cite{aflplusplus_argv}}
extends AFL to support command-line argument fuzzing by processing command-line options using fixed-length data chunks. 
\ts{The tool allows}{However, it applies} traditional file-based fuzzers \ts{to explore command-line argument spaces}{in a straightforward manner} without modifying the core AFL fuzzing algorithm.
TOFU~\cite{wang2020tofu} \ts{combines distance-guided target selection with structured input mutation for directed fuzzing. The tool}{} expands the fuzzing space to include command-line arguments and uses distance metrics to prioritize inputs that are closer to specified target basic blocks in the program. 
POWER~\cite{lee2022power} systematically explores command-line option configurations\ts{. The scheme first constructs}{ by constructing} diverse option configurations\ts{, selects}{ and selecting} maximally ``distant'' configurations based on function relevance\ts{, then performs file fuzzing with the selected option sets}{}.
However, these approaches lack robust validity checking \ts{mechanisms}{} for generated option combinations, resulting in many invalid configurations that lead to early program termination.

CrFuzz~\cite{song2020crfuzz} introduced clustering analysis to predict input validity\ts{ for multi-purpose programs}{}, demonstrating improvements in path and edge coverage\ts{ when integrated with state-of-the-art fuzzers like AFL}{}.
ConfigFuzz~\cite{zhang2023fuzzing} implements \ts{an innovative}{a} transformation technique that encodes program configurations as part of fuzzable input, allowing existing mutation operators to test program settings alongside normal inputs. 
CarpetFuzz~\cite{wang2023carpetfuzz} leverages natural language processing to automatically extract option constraints from documentation, significantly reducing the testing search space by filtering out invalid option combinations.
ZigZagFuzz~\cite{lee2025zigzagfuzz} improves fuzzing coverage by separately mutating command-line options and file inputs in an interleaving manner. 
However, these studies do not prioritize option combinations based on \ts{vulnerability likelihood}{their likelihood of improving coverage}, leading to inefficient exploration of the vast option space.

ProphetFuzz~\cite{wang2024prophetfuzz} uses LLMs to automatically predict high-risk command-line option combinations from program documentation. ProphetFuzz prioritizes combinations more likely to contain vulnerabilities and generates semantically coherent commands with appropriate input files.
However, ProphetFuzz has two fundamental limitations.
First, ProphetFuzz relies \ts{exclusively}{solely} on user manuals and online documentation to infer CLI options\ts{. This creates}{, causing} problems when documentation is incomplete, outdated, or insufficient for understanding option interactions.
Second, ProphetFuzz generates input files \ts{using}{through} generic Python scripts \ts{without tailored prompting}{rather than dedicated commands}, \ts{causing}{which leads } LLMs to produce \ts{arbitrary}{ad-hoc} files with limited \ts{}{validity and} variety.
In contrast, \sys analyzes source code\ts{, which provides ground truth about available options and their implementation, eliminating guesswork about what options exist and how they behave.}{ to obtain accurate option information and}
\ts{In addition, \sys employs an agentic sandboxed shell environment where LLMs can dynamically install tools, execute commands, and iteratively generate diverse input files tailored to each program's specific needs, achieving substantially higher coverage.}{ provides a sandboxed shell environment where the LLM can use real file-generation tools, enabling the generation of valid and diverse command-line options that reflect each program’s actual behavior and ultimately achieving substantially higher coverage.}

\noindent{\bf {Directed Fuzzing.}}
\textcolor{black}{Directed fuzzing such as AFLGo~\cite{bohme2017directed}, Beacon~\cite{huang2022beacon}, and SelectFuzz~\cite{luo2023selectfuzz} guide fuzzing toward specific target locations. 
AFLGo uses distance-based seed scheduling to prioritize inputs closer to target sites in the control-flow graph. Beacon uses lightweight static analysis to compute abstracted preconditions, enabling it to prune infeasible paths at runtime.}
SelectFuzz employs a precondition-guided approach that selects seeds capable of satisfying branch conditions leading to target locations, improving the efficiency of reaching specific code regions.
Other works such as Prospector~\cite{zhang2024prospector}, Liu et al.~\cite{liu2020large}, Cerebro~\cite{li2019cerebro}, She et al. ~\cite{she2022effective}, and Magneto~\cite{zhou2024magneto} have explored prioritization strategies and call chain decomposition for directed fuzzing.
However, these techniques target specific code locations through traditional input fuzzing and do not address the unique challenges of CLI programs where command-line option configurations significantly determine reachable program paths and vulnerability exposure.

\noindent{\bf {\textcolor{black}{Symbolic and Neuro-symbolic Execution.}}}
\textcolor{black}{
\sys's path-guided prompting is conceptually related to symbolic and concolic execution, including LLM-based variants such as ConcoLLMic~\cite{luo2026concollmic}, in that both use path information to guide input generation. However, \sys does not extract or solve path constraints. Instead, it relies on the LLM's semantic code understanding to identify required input formats and delegates generation to native tools that produce valid inputs by construction. These approaches are complementary: \sys generates semantically valid seeds, which constraint-based approaches can further explore.
}

\noindent{\bf {LLM-based Test Generation.}}
Numerous prior studies~\cite{yang2024enhancing, pizzorno2024coverup, dakhel2024effective, lemieux2023codamosa, xia2024fuzz4all} have proposed methods for \ts{test case generation}{generating test cases} using LLMs.
TELPA~\cite{yang2024enhancing} leverages program analysis to extract real usage scenarios and address hard-to-cover branches through feedback-based refinement.
CoverUp~\cite{pizzorno2024coverup} employs coverage metrics to iteratively guide LLMs toward improved line and branch coverage, significantly outperforming prior methods.
MuTAP~\cite{dakhel2024effective} utilizes mutation testing to identify weaknesses in generated tests and augment prompts with surviving mutants, enhancing bug detection capabilities. 
CITYWALK~\cite{zhang2025citywalk} specifically addresses C++'s complex features by incorporating project dependency analysis and language-specific knowledge to improve test correctness. 
These approaches demonstrate that combining structured program analysis with LLMs' generative capabilities yields the effective automated testing tools, 
\ts{however, these are unit testing methods and cannot be directly applied to CLI testing.}{but they focus on unit testing and operate at the function level, making them unsuited for exercising CLI behavior.}


Fuzz4All~\cite{xia2024fuzz4all} leverages LLMs as input generation and mutation engines to produce diverse, realistic inputs for multiple programming languages.
Oliinyk et al.~\cite{oliinyk2024fuzzing} presents research focused on embedded systems security, specifically targeting BusyBox, a ubiquitous software package in Linux-based embedded devices. 
However, these approaches rely on specific domain knowledge and cannot be universally applied to various programs.

\section{Limitations}

This section outlines the current limitations of \sys and highlights the issues that require further investigation.

First, we have not investigated the optimal number of iterations for iterative refinement (\autoref{subsec:iterative_refinement}).
We currently set $N_{trial}$ to 2, but this value may not be optimal for all programs and could depend on program complexity.
However, increasing the number of iterations is constrained by substantial token usage, which accumulates over multiple iterations.
This issue is exacerbated by \sys's current approach of retaining the entire chat history for iterative refinement.
Selective preservation of relevant context may be necessary to manage token costs efficiently and enable exploration of optimal iteration counts.

Second, we have not systematically evaluated \sys across different LLM models.
Our evaluation primarily uses Claude Sonnet 4.0 with limited comparison to GPT-4.1.
We observed that even among commercial LLMs, model selection significantly impacts performance, suggesting that smaller or open-source models may yield substantially different results.
Specifically, when using LLMs with limited reasoning capabilities and context windows, \sys would likely experience degraded performance.
Possible adaptations to support smaller models include decomposing prompts into smaller, focused subtasks that fit within limited context windows, using a pipeline architecture where separate LLM calls handle option extraction, file generation, and seed refinement independently, and employing retrieval-augmented generation (RAG) to provide relevant code snippets on-demand rather than loading entire codebases into context.
However, whether these adaptations can match the performance of large commercial models remains an open question.
Future work should systematically evaluate \sys across different LLM scales and architectures to understand the performance-resource trade-offs and identify minimum capability requirements.



\section{Conclusion}

In this paper, we presented \sys, a novel CLI fuzzing approach that leverages LLMs to generate semantically valid command-line option combinations and input files.
\sys combines path-guided context prompting, iterative refinement with coverage feedback, and autonomous orchestration of command-line tools for input file generation.
Our evaluation on 43 real-world programs shows that \sys discovered 51 previously unknown vulnerabilities, with 41 confirmed and 33 fixed by developers.
These results demonstrate that our approach establishes a foundation for more effective and automated LLM-based CLI fuzzing methods.

\section*{Acknowledgments}
We thank our anonymous reviewers, shepherd, and meta reviewer for their insightful feedback and guidance.
We also thank every member who helped in analyzing vulnerabilities. In particular, we thank Kosuke Fujioka, Akira Inoue, Yuto Inoue, and Gouki Iwasaki for their dedicated efforts.
This work was supported by JST, CREST Grant Number JPMJCR22M3, Japan.
In this research work, we used the UTokyo Azure (\url{https://utelecon.adm.u-tokyo.ac.jp/en/research_computing/utokyo_azure/}).





\bibliographystyle{IEEEtran}
\bibliography{references.bib} 

\clearpage

\appendices

\section{\textcolor{black}{Statistical Analysis Details}}\label{app:stats}
\textcolor{black}{
To assess whether the observed differences in coverage between \sys and each baseline are statistically significant, we apply the Mann-Whitney U test. We also report the Vargha-Delaney $\hat{A}_{12}$ effect size to quantify the magnitude of improvement. An $\hat{A}_{12}$ value of 0.5 indicates no difference, while values closer to 1.0 indicate that \sys consistently outperforms the baseline. Following standard conventions, we classify the effect size as negligible ($\hat{A}_{12} < 0.56$), small ($0.56 \leq \hat{A}_{12} < 0.64$), medium ($0.64 \leq \hat{A}_{12} < 0.71$), and large ($\hat{A}_{12} \geq 0.71$).
}

\textcolor{black}{
Table~\ref{tab:stat-all} presents the complete results for all four comparisons. \sys achieves statistically significant improvements ($p < 0.05$) on 95\% of programs against CarpetFuzz, 100\% against ProphetFuzz, 80\% against ZigZagFuzz, and 85\% against SelectFuzz. In the vast majority of significant cases, the effect size is large, confirming that the performance gains are not only statistically significant but also practically meaningful.
}



\begin{table*}[h!]
\centering
\caption{\textcolor{black}{Statistical comparison of each baseline vs.\ \sys (Mann-Whitney U test). $\Delta$\%: percentage improvement of \sys over the baseline. $\hat{A}_{12}$: Vargha-Delaney effect size. Sig.\ (significance level): * $p{<}0.05$, ** $p{<}0.01$, *** $p{<}0.001$.
Eff.\ (effect size category): L = large, M = medium, S = small, N = negligible.
$\uparrow$: higher is better; $\downarrow$: lower is better.}}
\label{tab:stat-all}
\scriptsize
\setlength{\tabcolsep}{3pt}
\begin{tabular}{l rrcrr rrcrr rrcrr rrcrr}
\toprule
 & \multicolumn{5}{c}{\textbf{CarpetFuzz vs.\ \sys}} & \multicolumn{5}{c}{\textbf{ProphetFuzz vs.\ \sys}} & \multicolumn{5}{c}{\textbf{ZigZagFuzz vs.\ \sys}} & \multicolumn{5}{c}{\textbf{SelectFuzz vs.\ \sys}} \\
\cmidrule(lr){2-6} \cmidrule(lr){7-11} \cmidrule(lr){12-16} \cmidrule(lr){17-21}
Program
 & $\Delta\%${\scriptsize$\uparrow$} & $p${\scriptsize$\downarrow$} & Sig. & $\hat{A}_{12}${\scriptsize$\uparrow$} & Eff.
 & $\Delta\%${\scriptsize$\uparrow$} & $p${\scriptsize$\downarrow$} & Sig. & $\hat{A}_{12}${\scriptsize$\uparrow$} & Eff.
 & $\Delta\%${\scriptsize$\uparrow$} & $p${\scriptsize$\downarrow$} & Sig. & $\hat{A}_{12}${\scriptsize$\uparrow$} & Eff.
 & $\Delta\%${\scriptsize$\uparrow$} & $p${\scriptsize$\downarrow$} & Sig. & $\hat{A}_{12}${\scriptsize$\uparrow$} & Eff. \\
\midrule
\midrule
avconv & $+177.6$ & $.0000$ & *** & $1.000$ & L & $+33.6$ & $.0000$ & *** & $0.875$ & L & $+61.6$ & $.0000$ & *** & $0.958$ & L & --- & $.0000$ & *** & $0.950$ & L \\
bison & $+14.8$ & $.0000$ & *** & $0.990$ & L & $+40.9$ & $.0000$ & *** & $1.000$ & L & $+1.2$ & $.4623$ &  & $0.510$ & N & $+48.4$ & $.0000$ & *** & $1.000$ & L \\
cflow & $+2.8$ & $.0000$ & *** & $1.000$ & L & $+29.1$ & $.0000$ & *** & $1.000$ & L & $+2.5$ & $.0083$ & ** & $0.723$ & L & $+24.7$ & $.0000$ & *** & $1.000$ & L \\
cjpeg & $+96.8$ & $.0000$ & *** & $1.000$ & L & $+28.0$ & $.0000$ & *** & $0.960$ & L & $+25.8$ & $.0013$ & ** & $0.780$ & L & $+497.0$ & $.0000$ & *** & $0.990$ & L \\
dwarfdump & $+16.9$ & $.0000$ & *** & $0.915$ & L & $+38.8$ & $.0000$ & *** & $1.000$ & L & $+15.3$ & $.0000$ & *** & $0.948$ & L & $+152.2$ & $.0000$ & *** & $1.000$ & L \\
ffmpeg & $+108.9$ & $.0000$ & *** & $1.000$ & L & $+60.8$ & $.0000$ & *** & $0.995$ & L & $+29.4$ & $.0007$ & *** & $0.795$ & L & $+1323.5$ & $.0000$ & *** & $1.000$ & L \\
gm & $+65.0$ & $.0000$ & *** & $0.990$ & L & $+65.6$ & $.0000$ & *** & $0.990$ & L & $+40.7$ & $.0001$ & *** & $0.845$ & L & $+556.7$ & $.0000$ & *** & $1.000$ & L \\
gs & $+13.1$ & $.0000$ & *** & $0.980$ & L & $+31.3$ & $.0000$ & *** & $1.000$ & L & $+0.9$ & $.4730$ &  & $0.507$ & N & $+257.0$ & $.0000$ & *** & $1.000$ & L \\
jasper & $+21.5$ & $.0000$ & *** & $1.000$ & L & --- & --- & --- & --- & --- & $+29.7$ & $.0000$ & *** & $0.922$ & L & $+4703.3$ & $.0000$ & *** & $0.950$ & L \\
mpg123 & $+39.9$ & $.0000$ & *** & $0.998$ & L & $+63.1$ & $.0000$ & *** & $1.000$ & L & $+3.9$ & $.0257$ & * & $0.681$ & M & $+48.2$ & $.5000$ &  & $0.501$ & N \\
nasm & $+9.5$ & $.0000$ & *** & $1.000$ & L & $+43.1$ & $.0000$ & *** & $1.000$ & L & $-30.8$ & $1.0000$ &  & $0.007$ & L & $+137.2$ & $.0000$ & *** & $1.000$ & L \\
objdump & $+66.7$ & $.0000$ & *** & $0.985$ & L & $+33.7$ & $.0000$ & *** & $0.900$ & L & $+70.0$ & $.0000$ & *** & $1.000$ & L & $+2547.9$ & $.0000$ & *** & $1.000$ & L \\
pspp & $+138.6$ & $.0000$ & *** & $1.000$ & L & --- & --- & --- & --- & --- & $+198.4$ & $.0000$ & *** & $1.000$ & L & $+139.4$ & $.0000$ & *** & $1.000$ & L \\
readelf & $+35.9$ & $.0000$ & *** & $0.875$ & L & $+50.2$ & $.0000$ & *** & $0.935$ & L & $+53.0$ & $.0001$ & *** & $0.850$ & L & $+54585.0$ & $.0000$ & *** & $1.000$ & L \\
tiff2pdf & $+51.5$ & $.0000$ & *** & $0.995$ & L & $+19.3$ & $.0000$ & *** & $0.875$ & L & $+29.2$ & $.0003$ & *** & $0.815$ & L & $+317.3$ & $.0000$ & *** & $0.865$ & L \\
tiff2ps & $+47.2$ & $.0000$ & *** & $0.970$ & L & $+16.8$ & $.0001$ & *** & $0.858$ & L & $+20.1$ & $.0000$ & *** & $0.920$ & L & $+111.4$ & $.0001$ & *** & $0.853$ & L \\
vim & $+11.9$ & $.1824$ &  & $0.585$ & S & $+65.0$ & $.0006$ & *** & $0.802$ & L & $+4.6$ & $.2804$ &  & $0.555$ & N & $-20.0$ & $1.0000$ &  & $0.035$ & L \\
xmllint & $+14.5$ & $.0001$ & *** & $0.855$ & L & $+48.0$ & $.0000$ & *** & $1.000$ & L & $+13.6$ & $.0169$ & * & $0.698$ & M & $+556.7$ & $.0002$ & *** & $0.825$ & L \\
xmlwf & $+5.4$ & $.0158$ & * & $0.700$ & M & $+14.4$ & $.0001$ & *** & $0.850$ & L & $+11.9$ & $.0016$ & ** & $0.774$ & L & $+5.6$ & $.2071$ &  & $0.576$ & S \\
yara & $+15.6$ & $.0000$ & *** & $1.000$ & L & $+60.7$ & $.0000$ & *** & $1.000$ & L & $+23.6$ & $.0000$ & *** & $0.970$ & L & $+241.8$ & $.0000$ & *** & $1.000$ & L \\
\bottomrule
\end{tabular}
\vspace{2pt}
\raggedright
\footnotesize \\ Eff.: L = large, M = medium, S = small, N = negligible. --- indicates no data available.
\end{table*}




\section{Correlation Analysis Details}\label{app:correlations}
This appendix provides detailed correlation analysis between graph structural properties and coverage performance for each target selection strategy discussed in Section~\ref{sec:rq4-2}. 
Table~\ref{tab:advantage_correlations} shows the complete correlation coefficients and p-values for all strategies across different graph metrics.

\begin{table}[h]
\centering
\caption{[RQ4-2] Significant correlations between graph structural features and relative branch coverage advantages over random selection.}
\label{tab:advantage_correlations}
\footnotesize
\begin{tabular}{llrr}
\toprule
\textbf{Strategy} & \textbf{Structural feature} & \textbf{Pearson $r$} & \textbf{$p$-value} \\
\midrule
CLOSE & diameter & 0.525 & 0.018 \\
CLOSE & avg\_shortest\_path & 0.472 & 0.036 \\
BET & pagerank\_top10\_concentration & 0.462 & 0.040 \\
BET & pagerank\_gini & 0.461 & 0.041 \\
DEG & closeness\_centrality\_skew & 0.457 & 0.043 \\
CLOSE & closeness\_centrality\_skew & 0.426 & 0.061 \\
CLOSE & largest\_scc\_size & -0.420 & 0.065 \\
DEG & pagerank\_top10\_concentration & 0.399 & 0.081 \\
CLOSE & largest\_scc\_ratio & -0.397 & 0.083 \\
DEG & pagerank\_gini & 0.392 & 0.087 \\
PAGE & largest\_scc\_size & -0.392 & 0.087 \\
DEG & diameter & 0.389 & 0.090 \\
PAGE & largest\_scc\_ratio & -0.388 & 0.091 \\
\bottomrule
\end{tabular}
\end{table}

\section{Decision Rule Generation}
\label{app:decision_rules}
Based on the significant correlations, we automatically generate decision rules for recommending strategies. For each strategy $s$ and significantly correlated feature $f$, we establish a threshold $\theta_f$ by computing the median value of $f$ across programs where $A(p,s) > 0$.

For positive correlations, we recommend strategy $s$ when $f(p) \geq \theta_f$; for negative correlations, when $f(p) \leq \theta_f$. Each decision rule is weighted by $|r|$ (absolute correlation). The overall confidence for recommending strategy $s$ to program $p$ is:
\begin{equation}
\text{Confidence}(p,s) = \frac{\sum_{i \in M_s} |r_i|}{\sum_{j \in R_s} |r_j|}
\end{equation}
where $M_s$ is the set of matched conditions and $R_s$ is all conditions for strategy $s$. Table~\ref{tab:decision_rules} summarizes the generated decision rules.
Table~\ref{tab:recommendations_summary} shows the distribution of chosen strategies for the 20 POWER dataset programs.

\begin{table}[!t]
\centering
\caption{Generated decision rules for strategy recommendation}
\label{tab:decision_rules}
\scriptsize
\begin{tabularx}{\columnwidth}{lXcrr}
\toprule
Strategy & Feature & Threshold & $|r|$ & Condition \\
\midrule
CLOSE & diameter & $\geq$ 10.0 & 0.525 & Large diameter \\
CLOSE & avg\_shortest\_path & $\geq$ 4.32 & 0.472 & Long paths \\
CLOSE & closeness\_skew & $\geq$ 5.22 & 0.426 & Skewed closeness \\
CLOSE & largest\_scc\_size & $\leq$ 3.0 & 0.420 & Small SCCs \\
CLOSE & largest\_scc\_ratio & $\leq$ 0.009 & 0.397 & Fragmented \\
\midrule
BET & pagerank\_top10\_conc. & $\geq$ 0.405 & 0.462 & Concentrated PR \\
BET & pagerank\_gini & $\geq$ 0.406 & 0.461 & High PR inequality \\
BET & pagerank\_skew & $\geq$ 8.18 & 0.376 & Skewed PR \\
BET & density & $\leq$ 0.003 & 0.375 & Sparse graph \\
BET & diameter & $\geq$ 10.0 & 0.353 & Large diameter \\
\midrule
DEG & closeness\_skew & $\geq$ 5.22 & 0.457 & Skewed closeness \\
DEG & pagerank\_top10\_conc. & $\geq$ 0.405 & 0.399 & Concentrated PR \\
DEG & pagerank\_gini & $\geq$ 0.406 & 0.392 & High PR inequality \\
DEG & diameter & $\geq$ 10.0 & 0.389 & Large diameter \\
DEG & pagerank\_skew & $\geq$ 8.18 & 0.317 & Skewed PR \\
\midrule
PAGE & largest\_scc\_size & $\leq$ 3.0 & 0.392 & Small SCCs \\
PAGE & largest\_scc\_ratio & $\leq$ 0.009 & 0.388 & Fragmented \\
\bottomrule
\end{tabularx}
\end{table}

\begin{table}[t]
\centering
\caption{Distribution of function selection strategies across POWER programs.}
\label{tab:recommendations_summary}
\small
\begin{tabular}{lrrr}
\toprule
\textbf{Strategy} & \textbf{Count} & \textbf{\%} & \textbf{Avg confidence} \\
\midrule
PAGE & 8 & 40.0\% & 0.88 \\
BET & 7 & 35.0\% & 0.80 \\
CLOSE & 4 & 20.0\% & 0.56 \\
random & 1 & 5.0\% & 0.77 \\
\midrule
Total & 20 & 100\% & 0.80 \\
\bottomrule
\end{tabular}
\end{table}

\subsection{Detailed Cost Analysis}
\label{app:cost_details}

Table~\ref{tab:token-usage} presents the complete token usage and API costs for each program in the POWER dataset.

\begin{table}[h]
\centering
\caption{[RQ5] Input and output token usage and Anthropic API costs for target programs.}
\label{tab:token-usage}
\footnotesize
\begin{tabular}{l|rrrr}
\toprule
\textbf{Program} & \textbf{\# Chats} & \textbf{Input} & \textbf{Output} & \textbf{Cost (USD)}\\
\midrule
avconv & 38 & 4,897,240 & 81,451 & \$15.91\\
bison & 40 & 5,064,940 & 121,440 & \$17.02\\
cflow & 34 & 4,188,222 & 102,827 & \$14.11\\
cjpeg & 33 & 4,260,867 & 72,569 & \$13.87\\
dwarfdump & 46 & 6,195,032 & 97,352 & \$20.05\\
ffmpeg & 40 & 5,193,340 & 100,943 & \$17.09\\
gm & 40 & 5,307,676 & 92,965 & \$17.32\\
gs & 42 & 5,497,537 & 73,344 & \$17.59\\
jasper & 34 & 4,251,739 & 84,518 & \$14.02\\
mpg123 & 44 & 5,687,286 & 83,373 & \$18.31\\
nasm & 29 & 3,200,466 & 60,541 & \$10.51\\
objdump & 40 & 5,056,497 & 114,108 & \$16.88\\
pspp & 27 & 3,252,209 & 69,922 & \$10.81\\
readelf & 39 & 4,892,200 & 99,998 & \$16.18\\
tiff2pdf & 24 & 2,898,758 & 120,321 & \$10.50\\
tiff2ps & 27 & 3,277,745 & 113,568 & \$11.54\\
vim & 40 & 4,640,882 & 77,075 & \$15.08\\
xmllint & 43 & 5,349,191 & 82,767 & \$17.29\\
xmlwf & 38 & 4,290,179 & 62,524 & \$13.81\\
yara & 38 & 4,664,417 & 93,444 & \$15.39\\
\midrule
\textbf{Average} & 36 & 4,603,321 & 90,252 & \$15.16\\
\bottomrule
\end{tabular}
\end{table}

\clearpage


\textcolor{black}{
\section{Meta-Review}
The following meta-review was prepared by the program committee for the 2026
IEEE Symposium on Security and Privacy (S\&P) as part of the review process as
detailed in the call for papers.
}

\textcolor{black}{
\subsection{Summary}
This paper presents PILOT, an LLM-based fuzzing framework designed to generate semantically valid seeds through path-guided prompting. Specifically, PILOT selects target functions based on centrality metrics and provides their call paths as context to the LLM. PILOT then iteratively refines seed generation using coverage feedback to produce effective command-line options while orchestrating native tools for input file generation. The evaluation shows improvements in both code coverage and bug detection over state of the art.
}

\textcolor{black}{
\subsection{Scientific Contributions}
\begin{itemize}
\item 3. Creates a New Tool to Enable Future Science.
\item 5. Identifies an Impactful Vulnerability.
\item 6. Provides a Valuable Step Forward in an Established Field.
\end{itemize}
}

\textcolor{black}{
\subsection{Reasons for Acceptance}
\begin{enumerate}
\item This paper creates a new too to enable future science. PILOT creates a seed generation framework that builds on static analysis, LLMs, and fuzzing. As a seed generator, it is highly compatible with other tools, including fuzzers.
\item The paper identifies impactful vulnerabilities. It reports the discovery of 51 zero-days across 24 open-source programs. This includes bugs not found by state-of-the-art tools and widely used programs, contributing towards securing the open-source software ecosystem sustainably.
\item The paper provides a valuable step forward in an established field. PILOT outlines how the next generation of seed generation can look, and it demonstrates how to include powerful LLMs, guided by call-graph paths and enriched by autonomous code exploration, to substantially improve bug finding capabilities.
\end{enumerate}
}

\subsection{Noteworthy Concerns}
None.

\end{document}